\documentclass[fleqn,usenatbib]{mnras}

\usepackage{graphicx}
\graphicspath{ {./Figures/} }

\usepackage[T1]{fontenc}
\usepackage{float}

\DeclareRobustCommand{\VAN}[3]{#2}
\let\VANthebibliography\thebibliography
\def\thebibliography{\DeclareRobustCommand{\VAN}[3]{##3}\VANthebibliography}

\usepackage{graphicx}	
\usepackage{amsmath}	
\usepackage{amssymb}	
\usepackage{color}
\usepackage[export]{adjustbox}
\usepackage{comment}
\usepackage{newtxtext,newtxmath}

\newcommand{\Feh}{\mbox{$\mbox{[Fe/H]}$}}
\newcommand{\Mh}{\mbox{$\mbox{[M/H]}$}}
\newcommand{\afe}{\mbox{$\mbox{[$\alpha$/Fe]}$}}
\newcommand{\mgfe}{\mbox{$\mbox{[Mg/Fe]}$}}
\newcommand{\mgfep}{\mbox{$\mbox{[MgFe]'}$}}

\title[Inferring Helium abundance from integrated spectra]{Inferring the Helium abundance of extragalactic Globular Clusters using Integrated Spectra}

\author[H. J. Leath et al.]{
H. J. Leath,$^{1,2,3}$\thanks{E-mail: \href{mailto:henry.leath@unige.ch}{henry.leath@unige.ch} }
M. A. Beasley,$^{3,4}$
A. Vazdekis$^{3,4}$, N. Salvador-Rusiñol$^{3,4}$ and A. Gvozdenko$^{5}$
\\
$^{1}$Département d’Astronomie, Université de Genève, Chemin Pegasi 51, CH-1290 Versoix, Suisse\\
$^{2}$Astrophysics Research Group, Faculty of Engineering and Physical Sciences, University of Surrey, Guildford, Surrey, GU2 7XH, United Kingdom\\
$^{3}$Instituto
de Astrofísica de Canarias, E-38200 La Laguna, Tenerife, Spain\\
$^{4}$Departamento de Astrofísica, Universidad de La Laguna, E-38206 La Laguna, Spain\\
$^{5}$Department of Astrophysics/IMAPP, Radboud University, PO Box 9010,
6500 GL, The Netherlands
}

\date{Accepted XXX. Received YYY; in original form ZZZ}

\pubyear{2021}

\begin{document}
\label{firstpage}
\pagerange{\pageref{firstpage}--\pageref{lastpage}}
\maketitle

\begin{abstract}
The leading method for the determination of relevant stellar population parameters of unresolved extragalactic Globular Clusters is through the study of their integrated spectroscopy, where Balmer line-strength indices are considered to be age sensitive. Previously, a splitting in the highly optimised spectral line-strength index H$\beta_o$ was observed in  a sample of Galactic globular clusters at all metallicities resulting in an apparent "upper branch" and "lower branch" of  globular clusters in the H$\beta_o$ -- [MgFe] diagram. This was suggested to be caused by the presence of hot Blue straggler stars (BSSs), resulting in an underestimation of 'spectroscopic' ages in the upper branch. Over a decade on, we look to re-evaluate these findings. We make use of new, large Galactic Globular Cluster integrated spectroscopy datasets. To produce a large, homogeneously combined sample we have considered a number of factors including the radial dependence of Balmer and metal lines. 
Using this new sample, in disagreement with previous work, we find the splitting in H$\beta_o$ only occurs at intermediate to high metallicities (\Mh $>-1$), and is not the result of an increased fraction of BSSs, but rather is due to an increased Helium abundance. We explore the possible impact of varying Helium on simple stellar population models to provide a theoretical basis for our hypothesis and then use the relationship between upper branch candidacy and enhanced Helium to predict the Helium content of three M31 clusters. We discuss what this can tell us about their mass and fraction of first generation stars.

\end{abstract}

\begin{keywords}
Galaxy: globular clusters: general -- stars: blue stragglers -- galaxies: individual: M31
\end{keywords}



\section{Introduction}
\defcitealias{cenarro2008}{C08}

Globular Clusters (GCs) are compact, tightly gravitionally-bound systems that are some of the oldest observed in the Milky Way (MW). They are found associated with most galaxies ($\mathrm{M}_\star > 10^6 \mathrm{M}_\odot$) and although they have been actively researched for well over a century \citep[see][]{Herschel1789}, their formation and evolution remains debated. 
It is possible to examine Galactic GCs (GGCs) using resolved stellar photometry  with deep colour-magnitude diagrams (CMDs) due to their close proximity. 
Deep HST/ACS CMDs are available for almost 70 MW GGCs (\citealt{sarajedini2007}).  
This has allowed the determination of two key parameters: age \citep[e.g.][]{deAngeli2005, meissner2006, Dotter2010, leaman2013, VandenBerg2013, Go2014, Milone2014, Ni2015, deBoer2016} and metallicity \citep[e.g.][]{Harris1996, mucciarelli2008, larsen2012}. These age and metallicity measurements have revealed at least two sub-populations of GCs in the MW, a presiding population of very old GGCs spanning a wide range of metallicities and a smaller, younger population of GGCs that shows an anti-correlation between age and metallicity \citep[e.g.][]{Mar_n_Franch_2009}. The very old population of GGCs can also be split into two further sub-populations by analysing the age-metallicity relation: a more populous metal-poor sub-population and a metal-rich sub-population. Several GGCs in the Local Group (LG) also have CMDs provided by the ACS and their ages are found to be coeval with the very old population of MW GGCs \citep{Wagner2017}. 
An additional parameter of significance is the Helium abundance, which has been shown to correlate with various parameters; Cluster mass \citep{Milone2014, milone2015helium, wagner-kaiser2017}, the red giant branch (RGB) bump \citep{cassisi1997, nataf2013, wagner-kaiser2017, Lagioia2018}, Carbon and Nitrogen abundance \citep{wagner-kaiser2017} etc. The Helium abundance has also been shown to be tightly connected to the multiple stellar populations scenario in clusters \citep[e.g.][]{Milone2018}. The Helium abundance is characterised by the Helium mass fraction which is primarily calculated for GGCs by fitting isochrones to their CMD. Again this method relies on the close proximity of GGCs.

Beyond the LG, for extra-Galactic globular clusters (EGCs) resolved spectroscopy and photometry and their subsequent deep CMDs are not available due to instrument limitations. Therefore, integrated spectroscopy, that takes the sum of all the light of the stars in the stellar population, is used to evaluate these parameters (see e.g. \citealt{Beasley2020}). This can be done by measuring Balmer and metal spectral line-strength indices and comparing them to simple stellar population (SSP) models \citep[e.g.][]{Nelson2011}; an effective method due to the Balmer lines sensitivity to the effective temperature ($T_\text{eff}$) of the Main Sequence Turn Off (MSTO) of the stellar population \citep[e.g.][]{Buzzoni1994}. 
However, the Balmer lines are also affected by stars
other than those at the MSTO, this includes stars on the horizontal branch and other hot populations.
For example, \citet{cervantes2008} found a splitting of the Balmer line measurements at a given metallicity for a sample of Milky Way GCs, giving younger \emph{apparent} spectroscopic ages for a group of identified GGCs.  It is thought this rejuvenation could be caused by non-canonical stellar evolutionary stages, specifically Blue Straggler Stars (BSSs) and Horizontal Branch (HB) stars. Both blue HB stars and BSSs show distinguished Balmer lines and have a high $T_\text{eff}$ with respect to the MSTO ($> 6500$K). A higher $T_\text{eff}$ of the MSTO corresponds to a younger population. Therefore, they could be capable of imitating a younger stellar population \citep[][]{schiavon2004, trager2005, Graves2008}. 

Previously, (\citealt{cenarro2008}, hereafter \citetalias{cenarro2008}) evaluated the effect of HB and BSS stars on the Balmer line measurements of GGCs. They were able to conclude that, at fixed metallicities, BSSs are primarily responsible for the variations seen in H$\beta$ for the integrated spectra of GGCs. This was due to the correlation seen between the specific frequency of BSSs and the metallicity. Over a decade on, we look to revisit this work, using more recent integrated spectra \citep{waggs, Kim2016} and HB morphology data \citep{torelli2019}. To anticipate the main conclusions of this paper, instead of explaining these increased Balmer line measurements by a increased fraction of BSSs, we relate these enhanced Balmer line measurements to enhanced Helium abundance. A GC with an enhanced Helium abundance has been shown to result in bluer (hotter) HB stars and will extend into the extreme HB at higher Helium abundances \citep{lee2005, Milone2014}. This effect has the potential to overcome metallicity, producing blue HB stars in metal-rich regimes. Due to the distinct Balmer lines of hot HB stars, an enhanced Helium abundance could therefore result in enhanced Balmer spectral line-strength measurements. Also, increased Helium abundance in GCs has been shown to effect the position (increased $T_\text{eff}$) of the MSTO \citep{valcarce2012}, once again possibly enhancing Balmer spectral line-strength measurements.
Finally, changing the Helium abundance at fixed metallicity affects iron abundance and the inferred age of the population.
The impact of changing $Y$ (Helium mass fraction) on the Balmer lines offers a way, in principle, to infer $Y$ from integrated spectroscopy.
Our results allow us to predict the Helium abundance for EGCs in M31 using integrated spectroscopy. Utilizing the relationship Helium abundance has with cluster mass and the ratio of multiple stellar populations, we are able to then provide predictions for these parameters.  

This paper is structured as follows: in Section \ref{sec:data}, the data we use is presented. Section \ref{sec:meth} presents the corrections made for radial velocities, spectral line-strength indices, stellar population models used to calculate spectroscopic ages and the smoothing of spectra and models. At the end of this section, we describe the process of creating a large homogeneous sample by combining our GGC integrated spectroscopy data. We then identify "rejuvenated" GGCs, creating an artificial upper and lower branch, and investigate the possible causes of the observed splitting in Section\,\ref{sec:a}. We conclude this section by predicting the Helium abundance, cluster mass and ratio of multiple stellar populations of several M31 EGCs. In Section \ref{sec:sum} we present a summary of our results and discuss their implications with some comments on possible further work.

\section{The Data}
\label{sec:data} 
\subsection{Integrated spectroscopy}

\defcitealias{waggs}{U17}
\defcitealias{S05}{S05}
\defcitealias{Kim2016}{K16}

\subsubsection{Galactic Globular Clusters}

We used three GGC datasets which were later combined to give a single large and homogeneous set of GGC data. 
The first GGC dataset was the WiFeS Atlas of Galactic Globular
cluster Spectra (WAGGS) \citep{waggs} which contains 64 MW GCs and 24
GCs found in the MW's surrounding satellite galaxies, 
with ages ranging from 20 Myr to 13 Gyr. Using the WiFeS
integral field spectrograph on the Australian National University (ANU) 2.3\,m telescope, the instrument provides a wide
wavelength coverage
($3270 - 9050$\,\AA) and high resolution, $R \sim \mathrm{0.8}$\,\AA\ full-width at half-maximum (FWHM),
spatially-resolved spectroscopy. The majority of the WAGGS data (65 GGCs)
is selected from \citet{sarajedini2007},
where this data has been supplemented with further GGCs to expand the age and chemical
composition range of the sample, adding both intermediate--old ($1\,\mathrm{Gyr}
< \mathrm{age} < 10\,\mathrm{Gyr}$) and young ($< 1\,\mathrm{Gyr}$) systems thereby providing a sample of GCs that represents the 
total, local population, rather than being a complete catalogue. There
are CMD age measurements available for the majority of GCs \citep[e.g.][]{meissner2006, VandenBerg2013, Go2014, Ni2015},
as well as HB Morphology data being readily available
\citep{Milone2014, torelli2019}.

The second GGC dataset was also used in \citetalias{cenarro2008} 
and contains the integrated optical spectra of
41 GGCs obtained using the Ritchey-Chretien (R-C) spectrograph mounted on the
4-m Blanco Telescope at the Cerro Tololo Inter-American Observatory. For each GGC, the integrated light was obtained within  1 core radius ($r_c$). Each spectrum covers the wavelength range $\sim 3350 - 6340$\,\AA\ with a FWHM of $\sim$ 3.1 \AA. More information is
available in the source paper (\citealt{S05}, this dataset is hereafter referred to as \citetalias{S05}). As this data was previously used in \citetalias{cenarro2008}, its use in parallel with the two further datasets allowed for direct comparisons with that study.
The third and final GGC dataset used the Intermediate Dispersion
Spectrograph (IDS) on the 2.5m Isaac Newton Telescope (INT) with the 235 camera, the EEV10 CCD detector and the R900V grating. It contains the
integrated spectra for 24 GGCs, where the intergrated light was
observed within varying regions of each GGC ($0.5, 1, 2, 3, 4\,r_c$), with a high resolution
of $\sim$ 2\,\AA\ FWHM covering a narrower wavelength range of 
$\sim 4000 - 5400$\,\AA. The narrow wavelength range still provides the
information needed to measure the appropriate spectral line-strength
indices in this work. We refer the reader to the source paper for further details on the data acquisition (\citealt{Kim2016}, this dataset is hereafter referred to as \citetalias{Kim2016}).

\subsubsection{\label{EGC}Extra-Galactic Globular Clusters}

Data of GCs present in M31 are also included in this study to allow for Helium abundance predictions of EGCs and to compare their integrated spectroscopy with GGCs. Provided by \citet{Nelson2011}, the data were obtained using the Hectospec multifiber spectograph on the 6.5m MMT Observatory at a resolution of 5\,\AA\ for a wavelength range of $3270$-$9200$\,\AA. The sample contains over 250 `old' GCs ($\ge 10$\,Gyr) with a high signal-to-noise ratio (S/N), where the median value is 75 per \AA \ at 5200 \AA, ranging from $\sim 8-300$\,\AA$^{-1}$. Considering this and the high number of EGCs in the sample, we made a S/N cut to remove the lower quality data. Spectra with a S/N value below the cut are not included, where each EGC target has multiple obtained spectra. The cut was performed at S/N\,$= 40$\,$\text{\AA}^{-1}$. A summary of each GGC and this EGC dataset is shown in Table\,\ref{tab:1}.

\begin{table*}
	\centering
	\begin{tabular}{lccccc} 
		\hline
		Dataset & Telescope & Instrument & Wavelength Coverage & Resolution & Number of GCs\\
		& & & [\AA] & [\AA]\\
		\hline
		WAGGS & ANU (2.3\,m) & WiFeS & 3270 -- 9050 & $\sim$0.8 & 86\\
		S05 & Blanco (4\,m) & R-C & 3350 -- 6340 & $\sim$3.1 & 41\\
		K16 & INT (2.5\,m) & IDS & 4000 -- 5400 & $\sim$2 & 24\\
		M31 & MMT (6.5\,m) & Hectospec & 3270 -- 9200 & $\sim$5 & 316\\
		\hline
	\end{tabular}
	\caption{\label{tab:1}A summary of the properties of each dataset used. Column one gives the name of each dataset, column two gives the telescope and column three the instrument used. Column four gives their respective wavelength coverage and column five gives their resolution in Angstroms FWHM. Finally, column six gives the number of GCs in each dataset (only old GCs for M31 data)}
\end{table*}

\section{Methodology}
\label{sec:meth}

\subsection{\label{prep}Preparation of spectra for index measurements}

Before the measurement of spectral line-strength indices, the integrated spectra of all datasets needed to be corrected for radial velocities (RVs) and smoothed to common resolutions.
The integrated spectra of \citetalias{S05} have already been corrected for RV but the remaining GGC and EGC datasets have not. The M31 data has RVs provided in \citet{Strader2011} which we make use of. For the WAGGS and \citetalias{Kim2016} data we calculate RVs using the \textsc{fxcor} task in \textsc{pyraf}. The GC spectra were cross-correlated with the appropriate Medium resolution INT Library of Empirical Spectra (MILES) SSP model templates~\citep{vazdekis2015}, corresponding to models that match the age and total metallicity of the cluster.
The three GGC datasets were all lowered to a common resolution to allow for meaningful comparisons. This was done via Gaussian smoothing,
where the input spectra are put through a Gaussian filter which
modifies the input signal via convolution with the one
dimensional Gaussian distribution, with a width determined
by the desired resolution. The wavelength coverage used is
limited by the \citetalias{Kim2016} data ($\sim$ 4000 -- 5400
\AA). Within this range the common resolution for all GGC data is 3.1 \AA \ FWHM, so all spectra are lowered to it.

\subsection{\label{indices}Age and metallicity sensitive line-strength indices}

We measured the line-strength indices of the spectra using \textsc{lector} \citep{vazdekis2011}.
\textsc{lector} is also capable of applying the required shift in wavelengths for each spectrum (due to RV) before the measurement of indices, where \textsc{lector} itself does not shift the spectrum but the wavelength limits of each spectral line-strength index. This is how we obtained the RV corrected spectral line-strength measurements. For all three of the spectroscopic datasets, some
of the GGCs have more than one spectrum available, where the data has
been collected on a different date. In this case the mean average of
the index measurements of each spectra was calculated to give a
single index measurement for each GGC.

The key age sensitive spectral lines in the optical range are the Balmer
lines (H$\beta$, H$\gamma$, H$\delta$) and are widely used for the estimation of the ages of unresolved stellar
systems as they are
sensitive to the effective temperature ($T_\text{eff}$) of the MSTO
of GCs \citep{Buzzoni1994}. In \citetalias{cenarro2008}, it is noted that a separation in GGC
measurements is seen in the optimised index H$\beta_o$. Developed by
\citet{cervantes2008}, it effectively minimises the dependency
on metallicity for H$\beta$, instead favouring its sensitivity to
age. Other age sensitive line-strength index options include H$\alpha$, H$\gamma$ and H$\delta$. 
H$\alpha$ lies outside of the wavelength range for both
the \citetalias{S05} and \citetalias{Kim2016} data, so was not taken into consideration.
Also, we did not use H$\gamma$ and H$\delta$ because these indices lead to significantly less orthogonal model grids in comparison to that with H$\beta$ (and therefore H$\beta_o$) and because of their higher sensitivity to \mgfe\ abundance ratio variations, making it more difficult to obtain a reliable spectroscopic age for the cluster. We use the metallicity-sensitive index \mgfep, which is a
combination of the indices Mgb, Fe5270 and Fe5335 that are defined
by \citet{trager1998}. Developed by
\citet{Thomas2003}, it provides an insensitivity to $\alpha$/Fe,
where $\alpha$ is referring to the $\alpha$-element abundance \citep[e.g.][]{vazdekis2015}.
This makes it a good tracer for the total metallicity of the
stellar population. 

\begin{figure}
    \centering
    \includegraphics[width=\columnwidth]{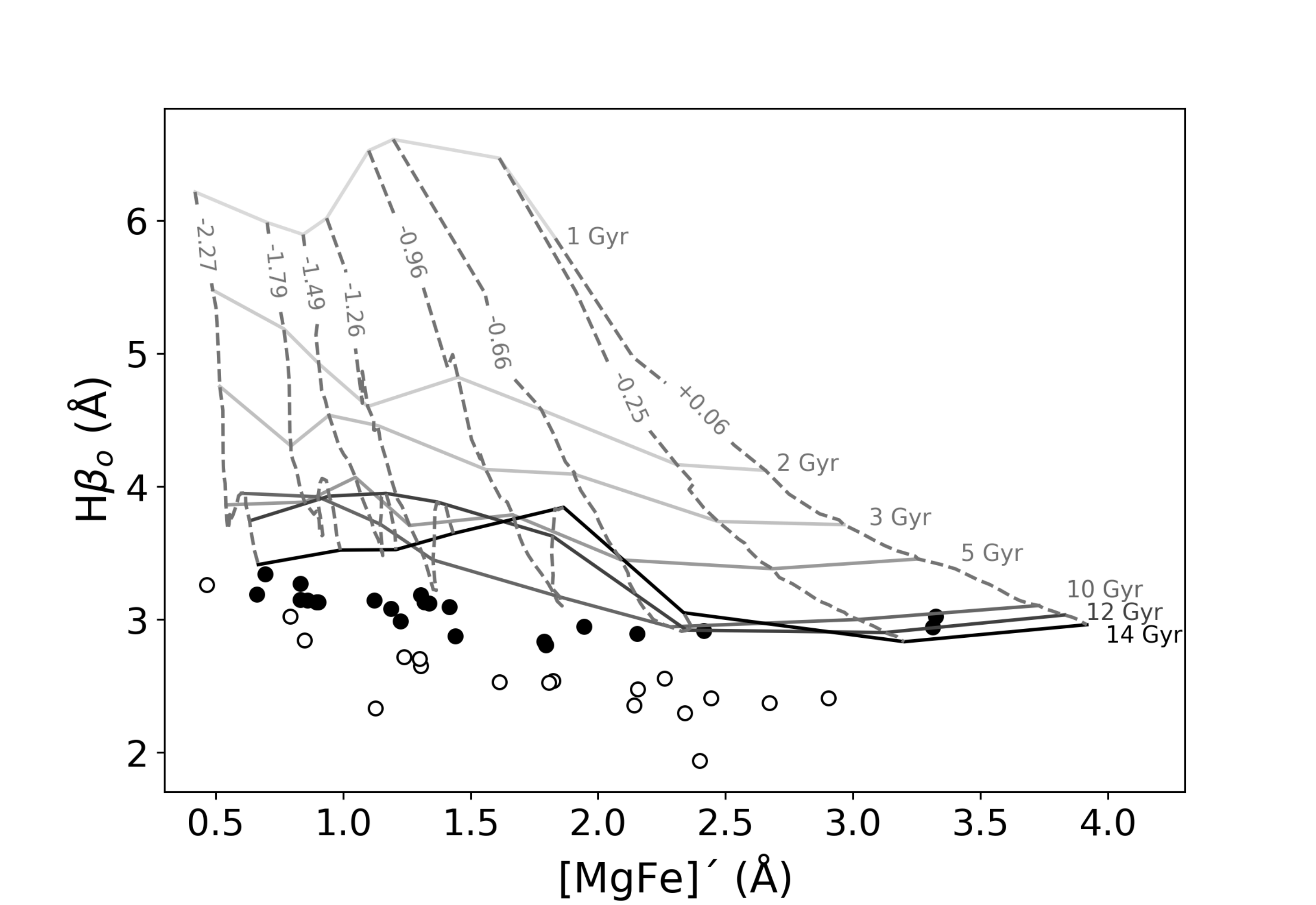}
    \caption{The spectral line-strength measurements of H$\beta_o$ against \mgfep\, for the \citetalias{S05} data. The upper and lower branch clusters, shown by black and white markers respectively, are selected in accordance with \citetalias{cenarro2008} making this figure a reproduction of their Fig.\,1a. The grey to black grid corresponds to the E-MILES SSP models \citep{EMILES}, the dashed lines are mono-metallic and the solid lines are coeval. The metallicity and age of each line are labelled accordingly.}
    \label{fig:1a}
\end{figure}

With both these indices measured, we are able to demonstrate the upper and lower branch splitting of H$\beta_o$ via the recreation of \citetalias[Fig.~1a]{cenarro2008}. We use the \citetalias{S05} data and the same upper and lower branch selection as \citetalias{cenarro2008} to produce Fig.\,\ref{fig:1a}, demonstrating the upper and lower branch splitting of H$\beta_o$.

\subsection{\label{SSP}Stellar population models}

To analyse the age and metallicity sensitive indices we use the
E-MILES SSP models
\citep{EMILES}. There are three main ingredients used for the
production of SSP models: A stellar library (either modelled or
empirical), a set of isochrones and an inference of the stellar
initial mass function (IMF). The E-MILES models solely use empirical
stellar libraries, spanning a wide range of wavelengths, starting at
the near-infrared (IRTF, \citealt{Cushing2005, Rayner2009}, CAT,
\citealt{cenarro2001}, Indo-US, \citealt{Valdes2004}) down to the
optical (MILES, \citealt{sanchez2006}) and finally to the UV (NGSL,
\citealt{Gregg2006}). These models are available for a range of IMF
shapes and slope values but we employ here a standard low-mass tapered "bimodal" IMF with logarithmic slope 1.30 for stars more massive than $0.6\,\mathrm{M_{\odot}}$.This IMF is close to that of the Kroupa Universal \citep{Kroupa2001}. The E-MILES
models use two sets of isochrones, where we use the BaSTI
scaled-solar theoretical isochrone models of \citet{Pietrinferni2004} converted to the observational plane on the basis of extensive photometric stellar libraries \citep[e.g.][]{Alonso1996,Alonso1999}.

These BaSTI-based models range from $0.03$ to $14$\,Gyr, which covers 
the range we require for our selection of GGCs. There are two GGCs
in the WAGGS data that are below this range but we are
only looking at the old GGCs in this data (> $10$\,Gyr). These models
also cover a range of metallicities $\Mh=-2.27$ to $+0.26$ where all the
GGCs used in this paper are comfortably found in this range (all 
are sub-solar). The empirical stellar spectra follow the MW abundance pattern with respect to \Feh. This gives models that are scaled-solar at solar metallicity but at lower metallicities lack consistency where scaled-solar isochrones are combined with $\alpha$-enhanced spectra. Therefore, we in some cases require the models described in \citet{vazdekis2015}, which cover the optical wavelength range and are computed for varying \mgfe\ with the aid of theoretical stellar spectra. The age and metallicity sensitive line strength indices of the SSP model synthetic spectra were measured and smoothed in accordance with the methodology laid out in Section\,\ref{indices}. The M31 integrated spectra data has a lower resolution of 5\,\AA, so the E-MILES models were smoothed to this resolution to allow for their subsequent comparison. For comparisons with the GGC integrated spectra, the SSP models were smoothed to the common resolution of 3.1\,\AA . From here, the three GGC datasets can now be meaningfully compared and combined, once we assess the role differing extraction window sizes play in variations of index values measured between our three GGC datasets.

\subsection{Producing a homogeneous sample}
\label{sec:samp}

\subsubsection{Observations}

Unlike the \citetalias{Kim2016} and \citetalias{S05} data, the WAGGS data was observed using a single, central pointing for each GC. This does not take into account the varied heliocentric distances of each GC (2.2 - 137 kpc), resulting in a substantial difference in the fraction of light observed.  
The mean radius of the WAGGS field-of-view (FoV) was 17.4 arcsec which encompasses between 0.12 (NGC\,5139) and 13 (Fornax 5) $r_c$ \citep{waggs}. This difference in observed $r_c$ is concerning due to at least three separate effects. Firstly, a reduced extraction window causes a reduced FoV mass being sampled. This can introduce stochastic effects where each stellar evolutionary stage is not evenly sampled \citep{cerv2013}. However, our use of the indices at the blue end of the spectrum ($\sim 5000$\AA) should help to negate these effects due to the increased stability in this region. This stability can be attributed to the lower intrinsic scatter seen at blue wavelengths compared to red wavelengths as a result of fewer stars contributing to the red wavelengths compared to the blue in relative and absolute terms \citep{cerv2013}. Also, as a result of the dynamical evolution of the clusters, the more massive stars sink to its centre while less massive stars are pushed to more external orbits. This results in the variation of the slope of the mass function with radius \citep[e.g.][]{andreuzzi2004, beccari2015, sollima2016}. Finally, it is known that the radial distribution of multiple stellar population is not constant \citep[e.g.][]{Larsen2015, Simioni2016, Nardiello2018}. These different populations have varying chemical abundances. Therefore, mean chemical abundances will also vary with radius.

\begin{figure}
    \centering
    \includegraphics[width=\columnwidth]{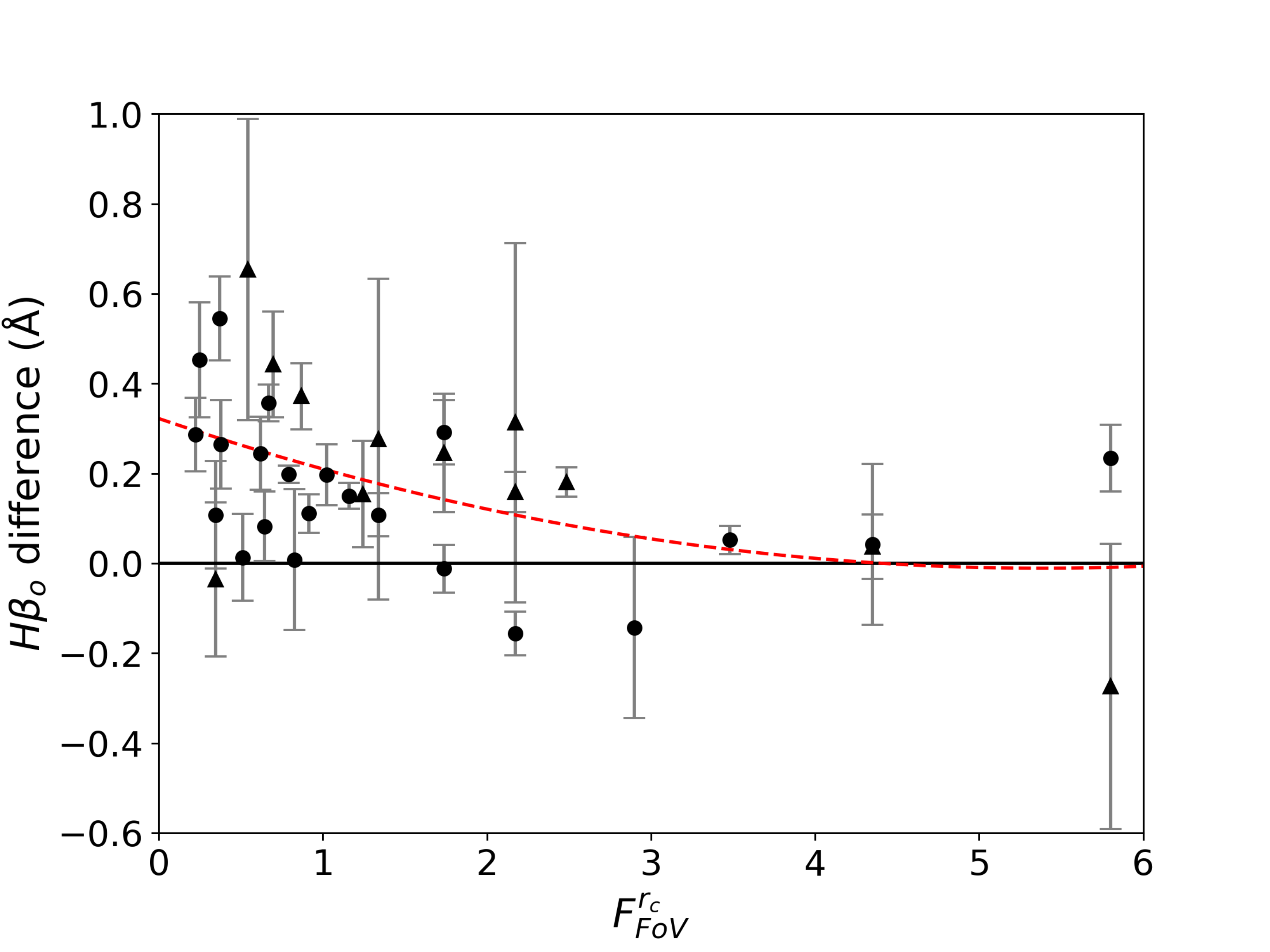}
    \caption{The H$\beta_o$ difference (WAGGS - \citetalias{S05}) as a function of the fraction of core radius, $F^{r_c}_{FoV}$. This shows a clear relationship between the two. We fit this relationship using a second order least squares polynomial, which is shown as the dashed red line. Errorbars for each point are shown in grey and are calculated with respect to the S/N of both the WAGGS and \citetalias{S05} spectra. Clusters with higher metallicity (> -0.66 dex) are shown as triangle markers while those with lower metallicities as circle markers.    }
    \label{fig:Hb_diff}
\end{figure}

To evaluate the effect this could be having on our data, we assessed the role that the fraction of $r_c$, $F^{r_c}_{FoV}$, observed has on Balmer line measurements by comparing the WAGGS data to our other datasets. The WAGGS data shares the most GC targets with the \citetalias{S05} sample, where the integrated spectra have all been measured at 1 $r_c$. In Fig.\,\ref{fig:Hb_diff} we compare the difference in H$\beta_o$ values between the two datasets with the $F^{r_c}_{FoV}$ (this data is provided in \citet{waggs}). We exclude NGC\,6362 from both the plot and the fit due to the anomalous nature of its high (> 1) H$\beta_o$ difference. Upon investigation, we find there to be an uncharacteristic spike in the WAGGS spectrum of NGC\,6362 at $\sim4904$\,\AA\,, which is within the wavelength region that defines H$\beta_o$'s red pseudo-continuum. Considering this we do not combine or compare the spectral line-strength measurements from the WAGGS and \citetalias{S05} data, leaving the two measurements separate. Note that the sensitivity to possible segregation effects is maximised for this index as it is mostly contributed to by hot MSTO stars, which are among the most massive stars that are alive in the stellar populations of the cluster.
 We see a clear and obvious relationship between the H$\beta_o$ difference and $F^{r_c}_{FoV}$, illustrated by the second order fit. Note also that the observed index differences for $F^{r_c}_{FoV}<2$ are larger than the typical errorbars of the cluster spectra, visible in Fig.\,\ref{fig:Hb_diff}. The errorbars were calculated with respect to the S/N of both the WAGGS and \citetalias{S05} data, where the WAGGS S/N for each spectra were provided by \citet{waggs} and the \citetalias{S05} S/N values were calculated from the auxilary information, multispectrum files made available. Individual errors were then calculated using the program \textsc{lector} \citep{vazdekis2011} which provides index error estimates on the basis of photon statistics under Poissonian consideration, taking into account the red, central and blue bandpasses of the specified spectral line-strength index. The errors from each dataset were combined in quadrature.

It is worth noting the majority of clusters with the greatest difference in H$\beta_o$ relative to their $F^{r_c}_{FoV}$ have higher metallicties (> -0.66 dex), shown as triangular markers in Fig.\,\ref{fig:Hb_diff}. However, when fitting separate relationships for higher and lower metallicity clusters in Fig.\,\ref{fig:Hb_diff} there was minimal difference for each fit (the error in each fit was greater than this difference). Hence, we continue with a single fit for all metallicities.

Considering the above discussion, we now looked to manipulate SSP models to determine whether or not mass segregation could be the cause, where its relationship with metallicity is also investigated.

\subsubsection{Mass segregation models}
\label{msmodel}
To further understand the effect mass segregation is having on the H$\beta_o$ index and how this relationship evolves with changing metallicity, we produce models that account for the change in the ratio of massive to lower mass stars. We use the base BaSTI-based models available from \citet{vazdekis2015} with a Kroupa IMF; Kroupa and bimodal IMF of 1.30 are very similar so this is acceptable for the intended comparisons \citep{EMILES}. To simulate the change in ratio of the massive to less massive stars in a GC we split the SSP models to create two partial SSPs (pSSPs), computed by integration along the isochrone from the lowest stellar mass up to a given stellar mass (pSSP$_\text{bottom}$), or from that mass up to the largest stellar mass that is alive in the stellar population (pSSP$_\text{top}$). We produce two sets of pSSPs; for the first they are cut just below the MSTO, already in the Main Sequence (MS) ($M_\text{cut} = 0.70M_\odot$) and for the second they are a cut at the base of the RGB. The latter cut varies as a function of metallicity: $M_\text{cut} = 0.80, \ 0.82, \ 0.87 \ \text{and} \ 0.934 \ M_\odot \ \text{for} \ \text{[Fe/H]} = -2.27, \ -1.26, \ -0.66 \ \text{and} \ -0.25 \ \text{dex}$. The pSSPs are then combined as shown in Eq. \ref{eq:pSSP}. Where $f_s$ is the so-called 'separation factor' and is used to adjust the ratio of massive to less massive stars in the models. We vary the separation factor between 1 (as a reference value that leads to the SSP) and 1.5. 

\begin{equation}
    \text{SSP} = \text{pSSP}_\text{bottom} + f_s\cdot\text{pSSP}_\text{top}
    \label{eq:pSSP}
\end{equation}

The \citetalias{Kim2016} data has the advantage of being available with various extraction windows ( 0.5, 1, 2, 3, 4 $r_c$). The various extraction windows allowed the direct comparison of the change in H$\beta_o$ between a) two extraction windows, $0.5 - 1 r_c$, (observations) and b) two ratios of massive to less massive stars, $\text{SSP}_{(f_s = 1.5)} - \text{SSP}_{(f_s = 1)}$, (models). 

\begin{figure}
    \centering
    \includegraphics[width=\columnwidth]{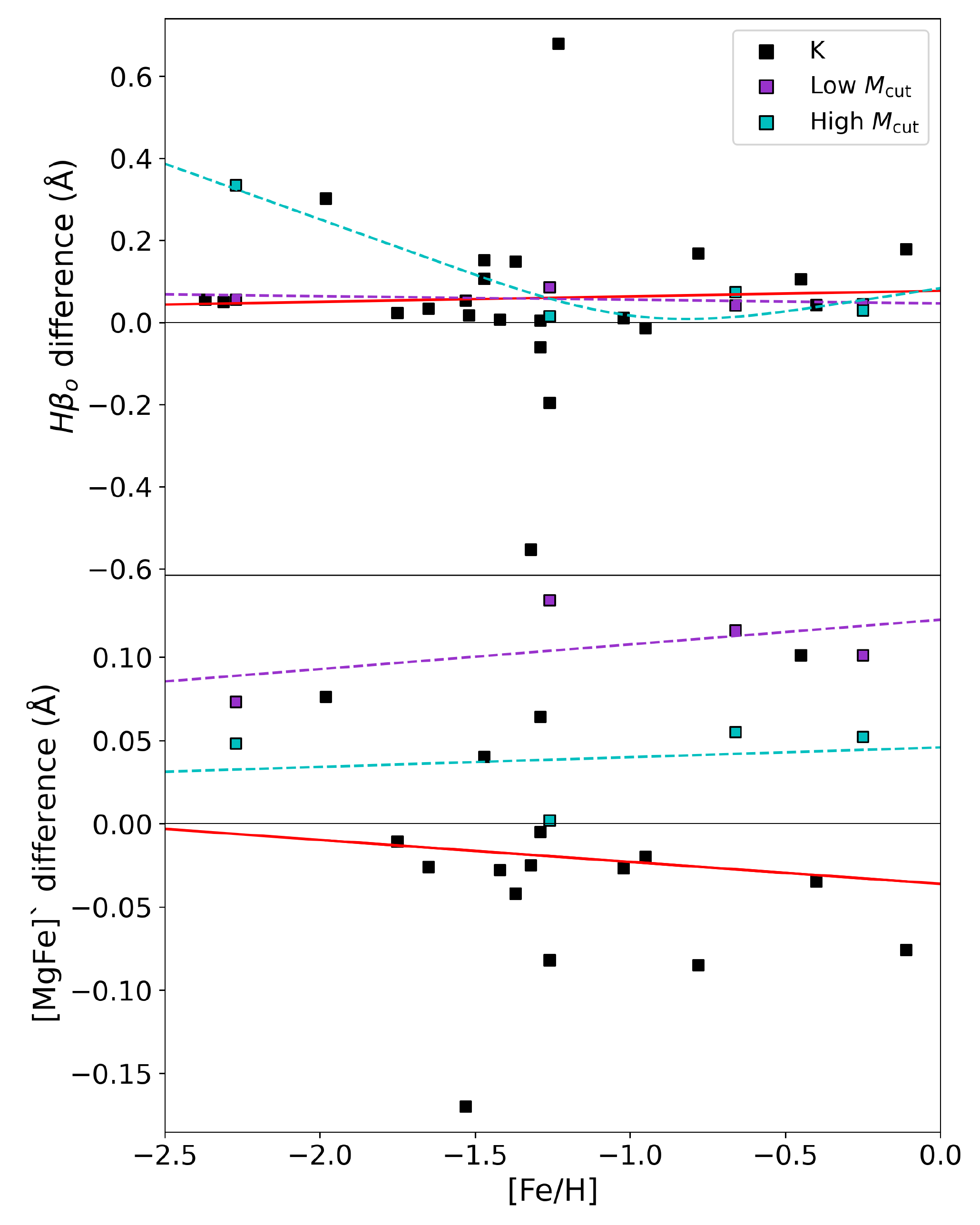}
    \caption{ The difference in H$\beta_o$(\textit{top}) and \mgfep(\textit{bottom}) between 0.5 and 1 $r_c$ extraction windows for the \citetalias{Kim2016} data as a function of metallicity are shown as the black square markers (labelled K). We fit these relationships using a first degree least squares polynomial, shown as a solid red line and of the form $y\,=\,(0.01335\pm0.08567)x\,+\,(0.07693\pm0.12201)$ and $y\,=\,(-0.01318\pm0.03358)x\,-\,(0.03619\pm0.04285)$ respectively. The difference in H$\beta_o$(\textit{top}) and \mgfep(\textit{bottom}) between separation factors of 1.5 and 1 for SSP models produced from pSSPs are shown as the cyan and purple square markers, they correspond to the high and low mass cut pSSPs respectively. We fit the relationships for each set; the dashed purple lines correspond to a first order polynomial fit of the low cut data points of the form $y\,=\,(-0.00877\pm0.01495)x\,+\,(0.04652\pm0.02012)$ and $y\,=\,(0.01484\pm0.01795)x\,+\,(0.1225\pm0.02415)$ respectively. The dashed cyan lines correspond to a natural cubic spline fit (\textit{top}) and a first order polynomial (\textit{bottom}) of the high cut data points.}
    \label{fig:pSSP}
\end{figure}

We choose to look at the H$\beta_o$ difference between extraction windows of 0.5 and 1 $r_c$. which corresponds to $F^{r_c}_{FoV}\,=\,0.5$. Looking at Fig.\,\ref{fig:Hb_diff}, it is clear to see that a large portion of the clusters have $F^{r_c}_{FoV}\,\sim\,0.5$, guiding our choice to consider the difference between extraction windows of 0.5 and 1 $r_c$. This is shown by the black markers in the top panel of Fig.\,\ref{fig:pSSP}.
For the models, we measure H$\beta_o$ for the various metallicities, cuts and separation factors. We then consider the H$\beta_o$ difference between a $f_s$ of 1.5 and 1 as a function of metallicitiy for the low and high $M_{cut}$ models, as shown by the purple and cyan markers in the top panel of Fig.\,\ref{fig:pSSP}. Comparing the observational and modelled fits, the observational fit agrees with the modelled first order fit for the lower cut partial SSPs to a reasonable degree when regarding the pure values of H$\beta_o$ differences. They appear to disagree with respect to the H$\beta_o$ difference's relationship with metallicity. However, the difference in slopes is insignificant when considering that both fits lie in the error bars of the other. The observations match somewhat with the predictions of the high cut models at higher metallicities but are in disagreement at lower metallicities. 

Hence, we can say our models predict that from 1 $r_c$ to 0.5 $r_c$ the ratio of massive ($> 0.70M_\odot$) to less massive ($< 0.70M_\odot$) stars increases by a factor of 1.5. However, this is only the case if mass segregation is the only cause of the observed H$\beta_o$ differences. On the bottom panel of Fig.\,\ref{fig:pSSP}, we look at \mgfep . We find that the \citetalias{Kim2016} data sees little significant variation with metallicity and a slight decrease in value with a decreasing radius. Focusing on the low cut pSSP models and their relation in the bottom panel as these models match H$\beta_o$ observations, we see that they disagree with respect to both the overall change in \mgfep\ with a decreased extraction window and how \mgfep\ changes with metallicity. Both disagreements are shown by the linear fits of the observed and lower cut models being outside their respective error bars. The disagreement between the low cut models and the observations for \mgfep\ either suggest that our pSSP models do not accurately produce the effect of mass segregation, or that mass segregation plays a minimal role in the observed correlation between H$\beta_o$ differences for clusters with extraction windows of differing radii. 
Even though we cannot be certain of the cause of the difference in H$\beta_o$ between datasets, we can still correct for it.

\subsubsection{\label{combine}Combining the data}

To correct the whole WAGGS sample, artificial H$\beta_o$ difference values were produced for clusters that are only in the WAGGS sample (not in \citetalias{S05}). This was performed by populating the fit shown in Fig.\,\ref{fig:Hb_diff} according to the $F^{r_c}_{FoV}$ of each cluster. The H$\beta_o$ difference of each GGC was then adjusted to remove its dependency on $F^{r_c}_{FoV}$, with this alteration being applied purely by changes in the WAGGS H$\beta_o$ measurements. Note that for $F^{r_c}_{FoV} > 6$ (Fornax\,5), due to the quadratic nature of second order polynomial fit, values would have been shifted drastically producing uncharacteristic results. To account for this, at $F^{r_c}_{FoV} > 6$, the model value at $F^{r_c}_{FoV} = 6$, H$\beta_o \ \text{difference} \approx 0$, was used as a constant, so no shift was applied.

As the \citetalias{S05} data were all measured at 1 $r_c$, we expected that the H$\beta_o \ \text{difference} \approx 0$ when $F^{r_c}_{FoV} = 0$. However, it is clearly visible that this is not the case. To investigate this jump between the datasets, we compared the difference in line-strength measurements for the indices CN1 and \mgfep\ which we find are not significantly altered with a change in $F^{r_c}_{FoV}$. For these cases, we found that there is still this jump in index measurements from the \citetalias{S05} to the WAGGS data, leading to the conclusion that a blanket increase needed to be applied for all \citetalias{S05} H$\beta_o$ and \mgfep\ values. Only a simple increase across all $F^{r_c}_{FoV}$ was required for the \mgfep\ values as their is no significant relationship between $F^{r_c}_{FoV}$ and \mgfep\,(see Fig.\ref{fig:pSSP}). For the H$\beta_o$ values, we took the value of the model at $F^{r_c}_{FoV}=\,1$, H$\beta_o$ difference\,$= 0.21$\,\AA, shifting all the \citetalias{S05} H$\beta_o$ measurements up by this value. As mentioned, the magnitude of the H$\beta_o$ difference does appear to have some dependency on metallicity. We attempted to account for this by again separating the clusters into two groups, high ($>\,-0.66$\,dex) and low ($<\,-0.66$\,dex) metallicity. We produced separate fits for both groups and manipulated the data accordingly. But, this had a minimal effect on the final shift applied to the WAGGS sample. Therefore, moving forward, the data was shifted utilizing the single fit that encompasses the whole metallicity range to reduce error. 

\begin{figure}
    \centering
    \includegraphics[width=\columnwidth]{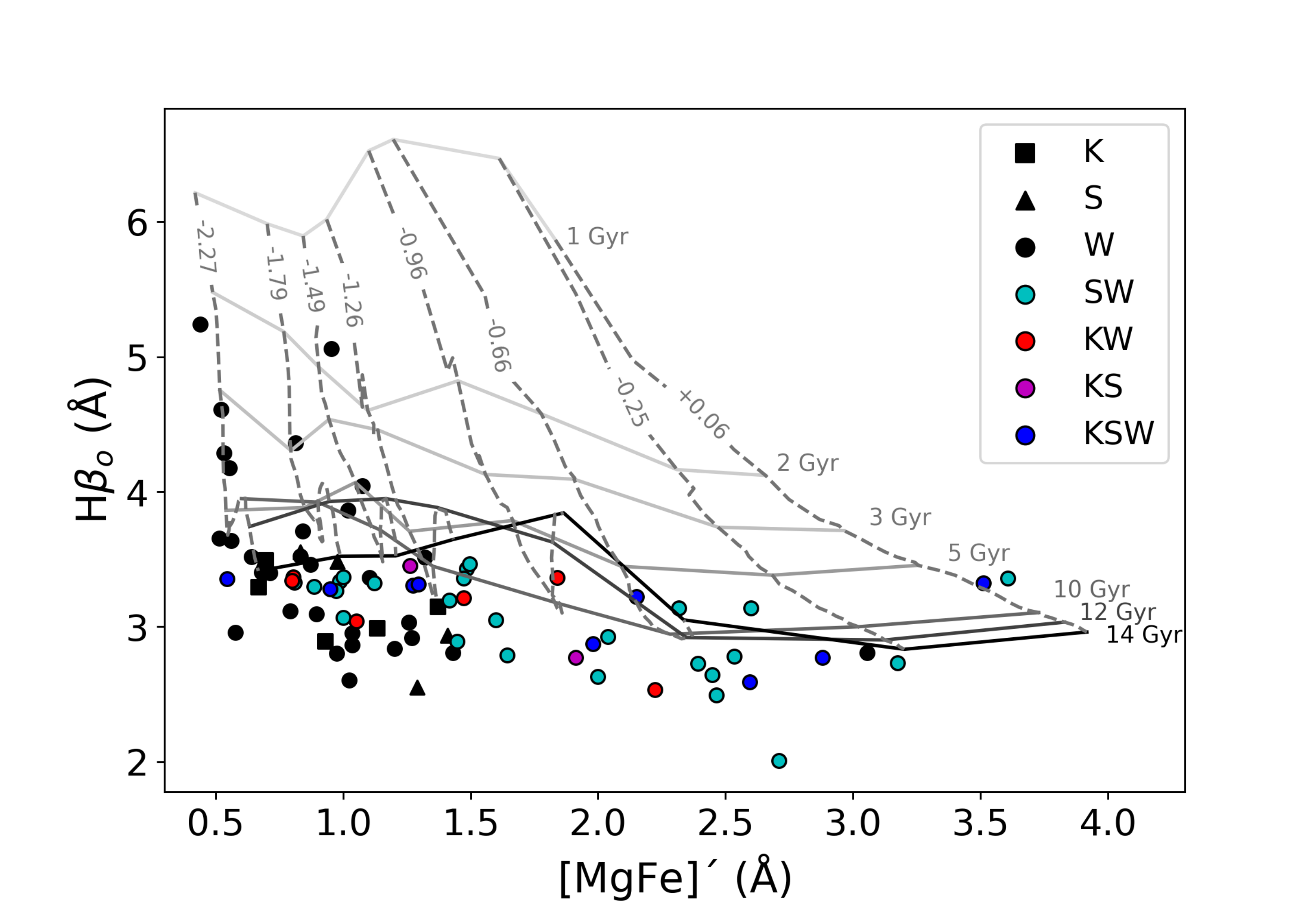}
    \caption{The spectral line-strength measurements of H$\beta_o$ against \mgfep\, for the combined dataset. The grey to black corresponds to the E-MILES SSP models \citep{EMILES}, the dashed lines are mono-metallic and the solid lines are coeval. Each line is labelled accordingly. Each colour and/or shape represents a different combination of datasets as described in the legend where W, S and K correspond to the WAGGS, \citetalias{S05} and \citetalias{Kim2016} data respectively. The mean error in H$\beta_o$ for each dataset combination is given in Table\,\ref{tab:2}.}
    \label{fig:new}
\end{figure}

With all three datasets agreeing, they were combined. The mean of the H$\beta_o$ and \mgfep\ measurements for each GGC available were taken, whether that is calculated from all three datsets, from two datasets or if only the measurement from a single dataset is available. This provided a large, homogeneous sample of 99 GGCs. The new dataset is shown in Fig.\,\ref{fig:new} where the age-sensitive index H$\beta_o$ is plotted against the metallicity-sensitive index \mgfep\ (only including old GGCs). The E-MILES models are then plotted over this data to help guide the eye and to show the predicted spectroscopic age and metallicity for each GGC.

\subsubsection{Uncertainties}
\label{uncertain}

Our new sample is only both meaningful and useful if the errors introduced through the method to fit the radial dependence of H$\beta_o$, the error characteristic of the spectra and the error caused by the combining of the data are not significant enough to effect upper and lower branch candidacy. The median S/N of the WAGGS data for the blue filter (4170 -- 5540\,\AA), which covers the wavelength range for H$\beta_o$ and \mgfep, is given as $\sim 77\text{\AA}^{-1}$ by \citet{waggs}. The \citetalias{S05} data provides the S/N of each pixel for each spectrum. The mean of the S/N was calculated over the wavelength range $4800 -- 4850$\,\AA\ to give a S/N value for each spectra, consistent with the methodology of \citet{waggs} (note the same methodology was used to calculate the S/N for the computation of errorbars in Fig.\,\ref{fig:Hb_diff}). The median of these S/N values was calculated to be $\sim 194\text{\AA}^{-1}$. Finally, the \citetalias{Kim2016} data has the sigma spectrum available for each spectrum. The standard signal spectrum was then divided by this sigma spectrum to give the S/N for each pixel. A median S/N value for this data was calculated using the same methodology used for the previous two datasets, with a value of $\sim 48\text{\AA}^{-1}$. 

The subsequent errors in the H$\beta_o$ measurements with respect to their S/N were  once again calculated using \textsc{lector}. For the WAGGS, \citetalias{S05} and \citetalias{Kim2016} data, errors were calculated as $\sim \pm0.15$, $\pm0.059$ and $\pm0.23$\,\AA\ respectively assuming a H$\beta_o$ value of $3.0$\,\AA. Error calculations are largely insensitive to the index value but this is still a reasonable selection with respect to the intermediate to high metallicity measurements shown in Fig.~\ref{fig:new}. Error propagation was used to consider the errors caused by the alterations applied to the WAGGS and the \citetalias{S05} data as described in Section\,\ref{combine}. The median value of $F^{r_c}_{FoV}$ $\sim 1.09$\,\AA\ was used as a part of error calculation. For the WAGGS and \citetalias{S05} data, post-shift errors were calculated and are presented in Table\,\ref{tab:2} (no shift was applied to the \citetalias{Kim2016} data).
The final H$\beta_o$ measurements were produced using a combination of the three datasets. Considering the aforementioned error values for each singular dataset, errors were calculated for each combination and are presented in Table\,\ref{tab:2}. It is also worth noting that spectral line-strength measurements of each GGC for all \citetalias{Kim2016} data, the majority of \citetalias{S05} data and some of the WAGGS data are the result of the combination of multiple spectral observations. Therefore, it is likely the error values presented here represent an overestimation of the true error values. The average dispersion between multiple measurements of individual clusters for the WAGGS, \citetalias{S05} and \citetalias{Kim2016} data are $0.120$, $0.097$ and $0.196$\,\AA \ respectively, suggesting the errors presented in Table\,\ref{tab:2} are slightly overestimated. Comparing all the dataset combinations and their respective errors to their position and the scale in Fig.~\ref{fig:new}, it is clear that they are not significant enough to effect a clusters upper and lower branch candidacy for the vast majority of GGC targets. The validity of the method used to combine the data is therefore confirmed.

\begin{table}
	\centering
	\begin{tabular}{lc} 
		\hline
		Combination & $\delta$H$\beta_o$\\
		& [\AA] \\
		\hline
		W & $\pm0.18$ \\
		S &  $\pm0.12$ \\
		K & $\pm0.23$ \\
		SW & $\pm0.11$ \\
		KW & $\pm0.15$ \\
		KS & $\pm0.13$ \\
		KSW & $\pm0.10$ \\
		
		\hline
	\end{tabular}
	\caption{\label{tab:2}The mean error for each combination of datasets shown in Fig.\,\ref{fig:new}. Column one gives the combination of datasets where W, S and K correspond to the WAGGS, \citetalias{S05} and \citetalias{Kim2016} data respectively. Column two gives the mean H$\beta_o$ error for each combination according to the methodology described in Section\,\ref{uncertain}.}
\end{table}

\section{The splitting of the age-sensitive H$\beta_o$ index}
\label{sec:a}

In this section we look to identify the upper and lower branch of clusters with relatively higher and lower H$\beta_o$ measurements, respectively, and first identified by \citetalias{cenarro2008}, for our combined data and M31 data using new methodology. Next, we investigate the possible origin of the apparent rejuvenation of those GGCs in the upper branch. Specifically, we consider the effect of  age, hot stellar evolutionary stages and the Helium abundance. We explore the effect of Helium on the H$\beta_o$ spectral line strength index via synthetic stellar spectra and isochrones. Then, we investigate the relationship between Helium abundance and two further cluster parameters: mass and the fraction of first generation stars. Finally, we use our findings to predict parameters in M31 upper branch clusters. 

\subsection{\label{rejuv}Identifying two branches}

\begin{figure*}
\centering
\centerline{\includegraphics[width=\linewidth]{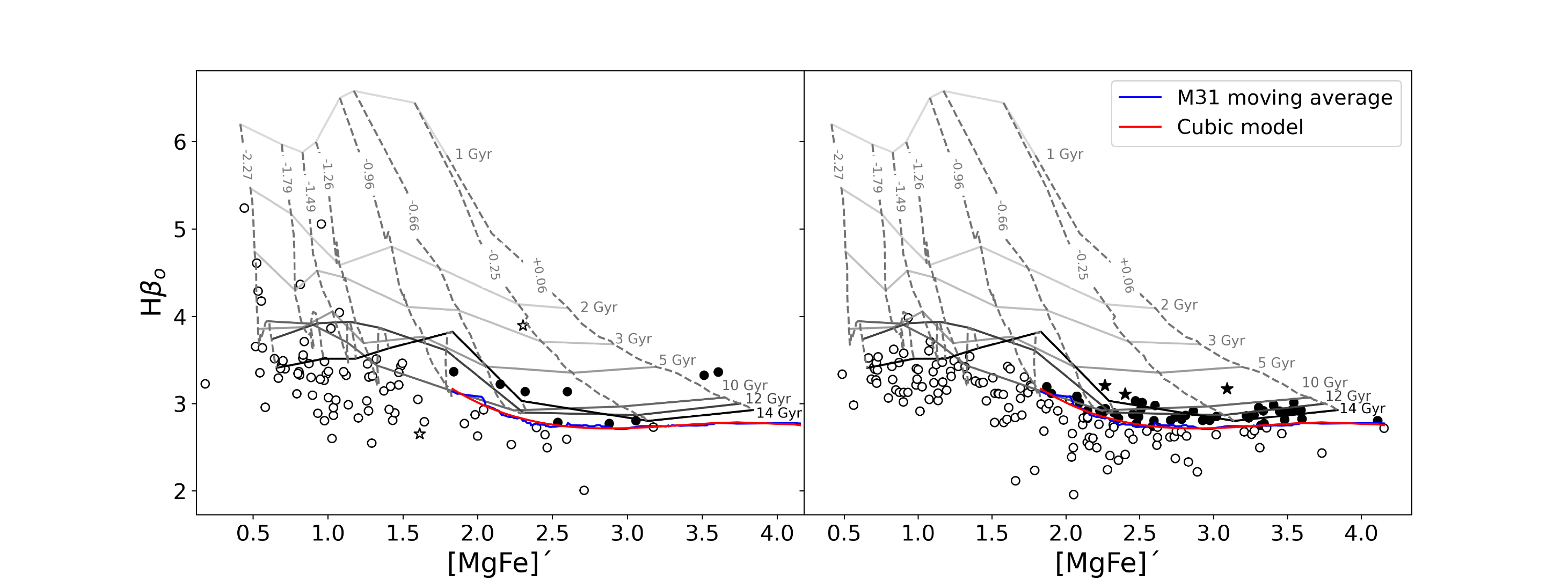}}
\caption{\label{fig:HL}The spectral line-strength
indices measurements of H$\beta_o$ for the combined data (\textit{left panel}) and the M31 data (\textit{right panel}). The grey to black grid corresponds
to the 
E-MILES SSP models \citep{EMILES}, with the ages and metallicities labelled.
The black solid circles represent the GGCs that are considered in the upper branch while the
open circles indicate those in the lower branch, as determined relative to the running mean of the M31 data shown by the solid blue line. We fit the moving average using a third order least squares polynomial given by the relation Eq.\,\ref{eq:model} and shown by the red line. On the left panel, the two star markers are the WAGGS and \citetalias{S05} measurements for the GGC NGC\,6362 which have not been combined and are not included in any further analysis. On the right panel, three upper branch clusters have closed star markers and are the subject of Helium estimates in Section\,\ref{sec:m31}.}
\end{figure*}

We look to identify two branches using
similar methodology to \citetalias{cenarro2008}, where GGCs
where shown to belong to either an upper or lower branch when using
the \citetalias{S05} data. Instead, we identify an upper and lower branch in our combined dataset and M31 dataset by first plotting the age-sensitive index
H$\beta_o$ against the total metalillcity-sensitive index \mgfep. The
E-MILES models are then plotted over this data to help guide the eye
and to show the predicted age and metallicity for each GGC. Most of
the GGCs lie below the model grid due to the well established
zero-point problem that effect SSP models
\citep[e.g.][]{gibson1999,vazdekis2001}, which has been suggested to be linked to atomic diffusion in stars near the MSTO and the enhancement of \afe\ abundance \citep{vazdekis2001}. However, our method does not depend on this model limitation as it relies on relative differences in positions of GGCs compared to each other, and to the model grid. Our selection is
demonstrated in Fig.\,\ref{fig:HL}, note that two separate
branches can only be identified at intermediate to higher
metallicities (\mgfep$ > 1.8$\AA ) compared to \citetalias{cenarro2008}
where they were able to identify both an upper and lower branch for
the full range of metallicities in the \citetalias{S05}.

Indeed there may not be two distinct branches at all, but rather a much larger spread in H$\beta_o$ than we would expect from the combined datasets (see e.g. Table\,\ref{tab:2}).
The upper branch (black markers) for both datasets is found above the running mean of the M31 data and the lower branch (white markers) beneath it. We have used the running mean of the M31 data as a selection tool as it is richer than our combined Milky Way GC dataset, especially at higher metallicities. We have fit the moving average using a third order least squares polynomial. Hence, we state that a GC is in the upper branch if it has line-strength index measurements of \mgfep$\,>1.8$\AA\ (\Feh$\,\gtrsim -1$) and follows the relation given in Eq.\,\ref{eq:model}. As demonstrated by the relatively small uncertainties in Eq.\,\ref{eq:model}, our fit provides a good estimation for the data.

\begin{align}\label{eq:model}
    \text{H}\beta_o > &(-0.169\pm0.010)\cdot\Feh^3 + (1.697\pm0.091)\cdot\Feh^2\nonumber \\&- (5.567\pm0.259)\cdot\Feh + (8.709\pm0.241)
\end{align}

Due to the availability of secondary data (e.g. CMD age, HB, BSS information) for GGCs and to allow for direct comparisons with \citetalias{cenarro2008} we first focus on the upper and lower branch of the combined dataset. 
The left panel of Fig.\,\ref{fig:HL}
presents our nine GGC upper branch candidates listed from lowest
to highest \mgfep: NGC\,6717, NGC\,6342, NGC\,6388, NGC\,6441, NGC\,6304, NGC6624, NGC\,6440, NGC\,6528 and NGC\,6553. Of these clusters, NGC\,6717 has an intermediate metallicity of [Fe/H]\,=\,-\,1.26 while the rest have high metallicities ($-\,0.55\,\leq$\,\Feh\,$\leq\,-\,0.11$). It is also worth noting the presence of 5 GGCs at lower metallicites with relatively high H$\beta_o$ values. These clusters do not fit the upper and lower branch shape presented in \citetalias{cenarro2008} so we exclude them from our upper branch selection.  

\begin{figure}
\centering
\centerline{\includegraphics[width=\columnwidth]{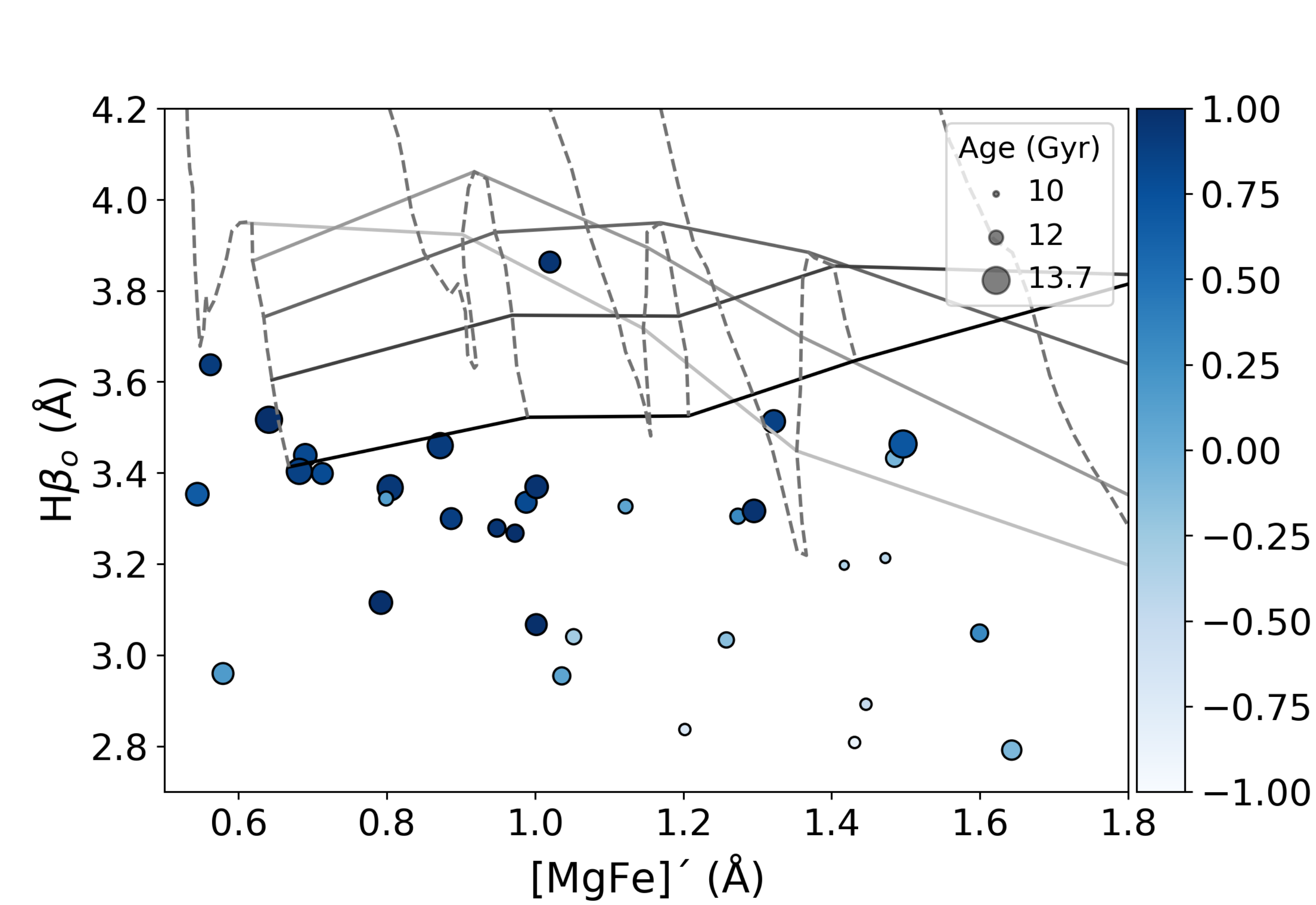}}
\caption{\label{fig:metal_poor}The spectral line-strength
indices measurements of H$\beta_o$ for the combined data as a function of \mgfep\,for metal poor clusters (\mgfep$< 1.8$\AA). The grey to black grid corresponds
to the 
E-MILES SSP models \citep{EMILES}, with the metallcities equal to that of Fig.\,\ref{fig:new}. The model age lines, from light to dark, correspond to $10, 11, 12, 13$ and $14$\,Gyr.
The size of each marker corresponds to the CMD derived age of each GC which range from $10$\,--\,$13.7\,$Gyr, as shown in the legend. The blue colour map indicates the $HBR$ of each cluster.}
\end{figure}

The lack of an upper and lower branch at low metallicities can be explained by taking into account a variety of parameters. It is first worth noting that at \mgfep$\,\lesssim\,1.8$\AA\ the relationship between metallicity and \mgfep\ experiences some degeneracy with age causing these clusters to bunch together as seen in Fig.\,\ref{fig:HL}. This relationship is also demonstrated by the increased concentration of the vertical metallicity SSP model lines at lower \mgfep. In fact it is possible to use a combination of age, metallicity and HB morphology to explain the previously observed splitting at low metallicities. In Fig.\,\ref{fig:metal_poor} we show how all three parameters are having an effect on the H$\beta_o$ measurements. The marker size indicates the age of the cluster ($10\le$~Age~$<13.7$\,Gyr), the blue colour map the HB morphology ($HBR$ index) and the vertical, dashed model lines the metallicity. Leaving apart the model zero-point issue the spread of the H$\beta_o$ values is rather similar to the spread predicted by the models at this age range where the horizontal, solid age model lines correspond to $10-14$\,Gyr taking steps of $1$\,Gyr (from light to dark). Therefore, a possible separation between the upper and lower branch is diluted within this expected scatter.

\subsection{Revisiting possible causes of splitting}

\begin{figure*}
    \centering
    \includegraphics[width=\linewidth]{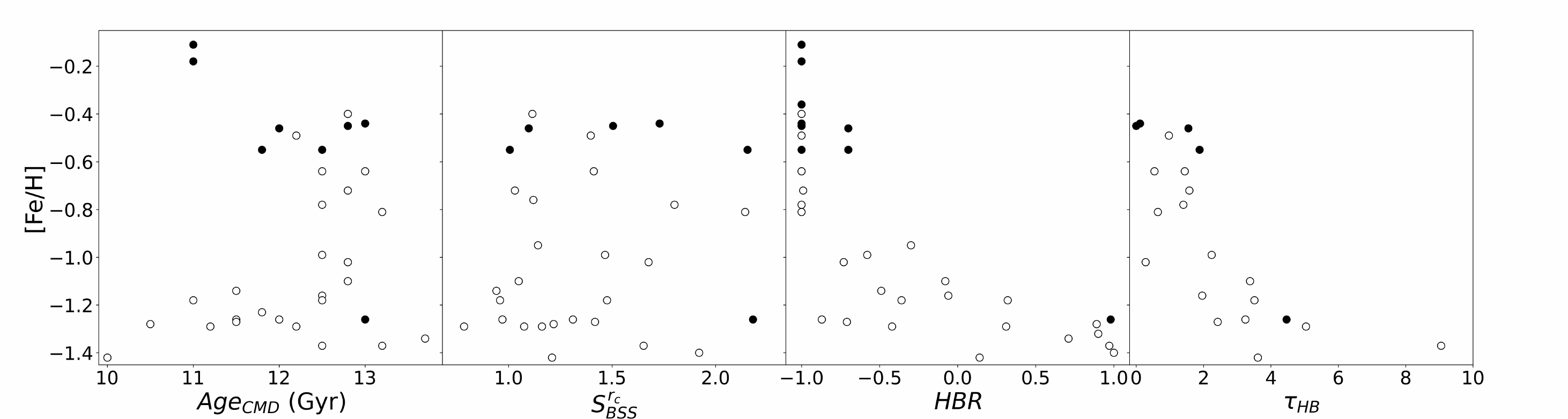}
    \caption{Counting from the left; \emph{First panel}: The metallicity of each GGC shown as a function of its age (taken from \citet{waggs}). \emph{Second panel}: The metallicity of each GGC shown as a function of its specific BSS fraction, $S^{r_c}_\text{BSS}$, with the data taken from \citet{moretti2008}. \emph{Third panel}: The metallicity of each GGC shown as a function of the HB morphology index $HBR$, mostly provided by \citet{Harris1996} where the data for NGC\,6388 is taken from \citet{Zoccali2000} and NGC\,6441 is assumed the same \citep{Puzia2002}. \emph{Fourth panel}: The metallicity of each GGC shown as a function of the HB morphology index $\tau_{HB}$, provided by \citet{torelli2019}. All metallicity data is provided by \citet{waggs}. Black markers correspond to GGCs assigned to the upper branch and white marker to the lower. }
    \label{fig:2e}
\end{figure*}

In this section we describe how we compare GGCs in the upper branch
to those in the lower branch at similar metallicities. This
comparison involves first looking at multiple properties of each stellar
population (CMD age, BSS fraction and HB Morphology) which were all considered by \citetalias{cenarro2008}. First 
we consider each GGCs age, calculated from their resolved CMD, and compare to the model grid in a relative sense to see if this is able to explain their presence in the upper branch. CMD ages are provided for the vast
majority of GGCs in \citet{waggs} from various
sources \citep[e.g.][]{deAngeli2005, meissner2006, Dotter2010,  Go2014, Milone2014, Ni2015, deBoer2016}. 
We look at two of our upper branch candidates, NGC\,6717 and NGC\,6342, with ages $13.0$\,Gyr and $12.5$\,Gyr, respectively. The SSP models for these
ages show a clear peak in H$\beta_o$ values at \mgfep$\sim2.4$\AA\ where
these GGCs are found. This is shown in Fig.\,\ref{fig:HL}
where the two darkest SSP model age lines,
corresponding to $12$ and $14$\,Gyr\footnote{The $12.5$ and $13$\,Gyr model lines do not deviate significantly from the two model lines shown. Therefore, we are able to use them comparatively.}, tend to rise in H$\beta_o$ at lower
metallicities (\mgfep$< 2.4$\AA). Taking into consideration the zero-point issue we see that these two clusters have higher H$\beta_o$ values than expected with respect to the models for these old CMD ages. Furthermore, we compare the aforementioned pair of upper branch clusters to two clusters at similar \mgfep\ and with similar or equivalent ages: NGC\,6171 (12.8\,Gyr) and NGC\,6838 (12.5\,Gyr) respectively. Both have H$\beta_o\sim2.5$\AA\ placing them in the lower branch, where their similar ages and metallicities (but significantly lower H$\beta_o$ than NGC\,6717 and NGC\,6342) in unison with the expected H$\beta_o$ values derived from the SSP models age lines suggests age alone cannot be used to explain the observed difference in the strength of H$\beta_o$ with respect to the lower branch for these clusters. The CMD ages of two upper branch candidates with higher metallicities, NGC\,6388 and
NGC\,6441, are of interest as they have a slightly lower age ($\le12$\,Gyr) than the
GGCs at similar metallicites in the lower branch (e.g. NGC\,6652\,$13.0$\,Gyr, NGC\,6838\,$12.5$\,Gyr, and NGC\,6637\,$12.5$\,Gyr). In this case Fig.\,\ref{fig:HL} shows that such a large jump in H$\beta_o$ strength cannot be attributed to the estimated difference in CMD age. To assess the effect of age on the splitting of H$\beta_o$ across all metallicities we recreate \citetalias[Fig.~2]{cenarro2008}. We first compare the age of each cluster to its metallicity, which is again provided by
\citet{waggs}, from various sources
\citep[e.g.][]{Harris1996, mucciarelli2008, larsen2012}. This can be seen in the first panel of Fig.\,\ref{fig:2e}. There are 5 upper branch GGCs that have lower ages compared to other clusters at similar metallicites (including NGC\,6388 and NGC\,6441). There is one cluster that appears to have a more average age of $12.5$\,Gyr (NGC\,6342) and then two with higher ages of $12.8$\,Gyr and $13.0$\,Gyr (NGC\,6304 and NGC\,6717). We have already discussed the role of the age in NGC\,6342 and NGC\,6717's presence in the upper branch. Looking at NGC\,6304, the age model lines in Fig.\,\ref{fig:HL} at this \mgfep\ value suggest that age has a minimal effect on its H$\beta_o$ value ($12$ and $14$\,Gyr lines crossover). It is worth noting that NGC\,6440 is excluded from the plot due to a lack of age measurement but is believed to be approximately coeval with NGC\,104 (\citealt{Origlia2008}, 47\,Tucane\,$12.8$\,Gyr) .

We now look to consider the effect of hot, non-canonical stars, specifically BSSs and hot HB stars.
In \citetalias{cenarro2008}, BSS data provided by \citet{moretti2008} were used exclusively,
providing a catalogue of BSSs extracted from a homogeneous sample of
56 GGCS. Using this data, the specific fraction of BSSs was
calculated, given by the logarithm of the number of
BSSs within 1 $r_c$, $N_\text{BSS}$, over the sample luminosity in
units of $10^4 \text{L}_\odot$ in the F555W Hubble Space Telescope band in the same aperture, $L_\text{F555W}$. 
This specific fraction is hereafter referred to as $S^{r_c}_\text{BSS}$ and directly uses the luminosity of a cluster to normalise the number of BSSs, parametersing their contribution to the integrated light of each GGC. The BSS data provided by \citet{moretti2008} remains the largest, homogeneous BSS sample available for our GGC data.
We us the BSS data and the combined GGC data, comparing $S^{r_c}_\text{BSS}$ to each clusters metallicity.
This is shown in the second panel of Fig.\,\ref{fig:2e}. In \citetalias{cenarro2008} they showed a clear
separation into two groups, where this matched the upper and lower
branch identified spectroscopically. However, this is not seen in
the second panel of Fig.\,\ref{fig:2e} where there is no clear separation in accordance with our selected upper and lower branch
groups. In fact, at the highest metallicities, upper branch candidates with the whole range of $S^{r_c}_\text{BSS}$ are clearly visible. This is likely due to the lack of a clear upper and lower branch separation at low metallicities, where these upper branch GGCs helped drive the correlation seen in \citetalias{cenarro2008}. The correlation in \citetalias{cenarro2008} in Fig.\,2e is the result of seven GGCs: NGC\,1851, NGC\,5904, NGC\,6171, NGC\,6266, NGC\,6284, NGC\,6342 and NGC\,6652. Of these seven, one (NGC\,6342) is in our identified upper branch. This is because NGC\,5904, NGC\,1851, NGC\,6284 and NGC\,6266 are at lower metallicities (\mgfep$< 1.6$) when compared to any of our upper branch candidates. NGC\,6652 does not have a suitably high H$\beta_o$ value in our combined dataset. So this correlation is largely dependent on the presence of an upper branch at low metallicities, in contention with what we have observed. 
This plot is held back by missing BSS data for three upper branch candidates: NGC\,6440, NGC\,6528 and NGC\,6553. Also in \citetalias{cenarro2008}, NGC\,6388 and NGC\,6441 are not considered even though they appear in the upper branch. This is due to their status as 'second parameter' clusters \citep[e.g.][]{rich1997}, with their high H$\beta_o$ values explained by the presence of hot HB stars.

Finally, we consider the role HB morphology plays in the splitting of H$\beta_o$. In
this paper the first index used to parameterise the HB is the
classic index $HBR$, first developed by \citet{Lee1994}. This index
is widely used due to its easy estimation from both theoretical and
observational perspectives. However, it suffers from a saturation in
both metal-rich and metal-poor regimes due to it simply being
defined as the fraction of the difference between the number of blue
and red stars, not taking into account the exact positions of all
stars along the HB. More recently, \citet{torelli2019} present a new
index, $\tau_{HB}$, which is defined as the area subtended by the
cumulative number distribution along the observed HB in
magnitude divided by the same in colour. This index is effective in
eliminating the saturation that hinders the $HBR$ index and is
therefore used in this paper. $HBR$ is used alongside $\tau_{HB}$
due to its wide availability, data is available for almost all required GGCs from \citet{Harris1996}. We then supplement this with $HBR$ data for the two metal-rich bulge clusters NGC\,6388 and NGC\,6441. The cluster NGC\,6338 has a $HBR$ of -\,0.70 taken from \citet{Zoccali2000} and NGC\,6441 is assumed to equivalent $HBR$ due to their similar HB morphologies \citep{Puzia2002}, as in \citetalias{cenarro2008}.
First using the parameter $HBR$ to characterize the HB morphology, we compare the $HBR$ value of each GGC to their metallicity, as shown in the third panel of Fig.\,\ref{fig:2e}. We have $HBR$ data available for all nine of our upper branch GGCs. The eight at high metallicities demonstrate the degeneracy of this index at both high and low metallicities as it is simply the fraction of the difference between the number of blue and red stars, so does not take into account the exact positions of all the stars along the HB. Because of this it is hard to draw any conclusions using this data at these metallicities. However, NGC\,6717 (intermediate metallicity) has a high $HBR$ of +0.98 compared to other clusters at similar metallicities, corresponding to a bluer HB. 
To combat the degeneracy of $HBR$, we use the new index $\tau_{HB}$ where this is shown as a function of metallicity in the fourth and final panel of Fig.\,\ref{fig:2e}. Using this parameter, we are unable to see any correlation between HB morphology and upper branch status. At higher metallicities, both upper and lower branch candidates occupy a similar range of $\tau_{HB}$ values ($0 \le$ $\tau_{HB}$ $\le 2$). This comparison is limited by the number of lower branch candidates at the same metallicity, which is perhaps a refection of the metal-rich nature of the majority of our upper branch candidates. Using this new index, NGC\,6717 still has a higher $\tau_{HB}$ value (bluer HB) than the majority of GGCs at similar metallicities but not to the extent shown when using $HBR$.  

Now we consider a new parameter that is likely causing the splitting at intermediate to high metallicities.

\subsection{Helium}

\begin{figure}
\centerline{\includegraphics[width=.93\columnwidth,left]{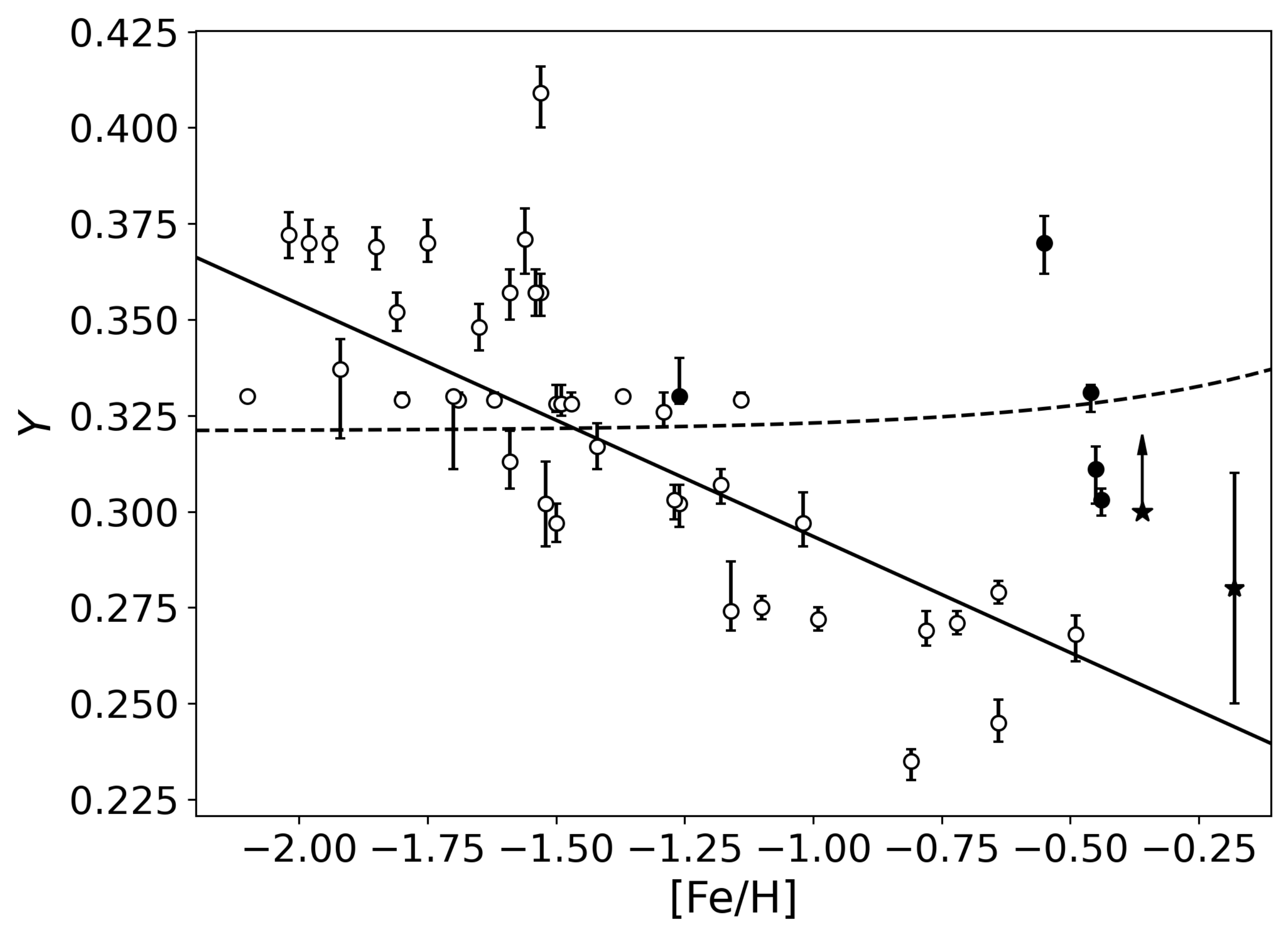}}
\caption{\label{fig:He1} The Helium abundance, $Y$, plotted as a function of metallicity for Galactic globular clusters. The solid black line represents the first order polynomial fit for lower branch clusters of the form $Y = (-0.0564\pm0.0084)\log(\Feh) + (0.24\pm0.013)$ against metallicity. The Helium abundance values and errors are provided by \citet{wagner-kaiser2017}. The two clusters marked as closed stars, NGC\,6440 and NGC\,6553, have data and errors (only NGC\,6553) provided by \citet{mauro2012} and \citet{Guarnieri1998} respectively. Black markers correspond to upper branch candidates and white to lower. The dashed line shows the relationship $\Delta Y/\Delta Z = 1.54$}. 
\end{figure}

\begin{figure}
\centerline{\includegraphics[width=.93\columnwidth,left]{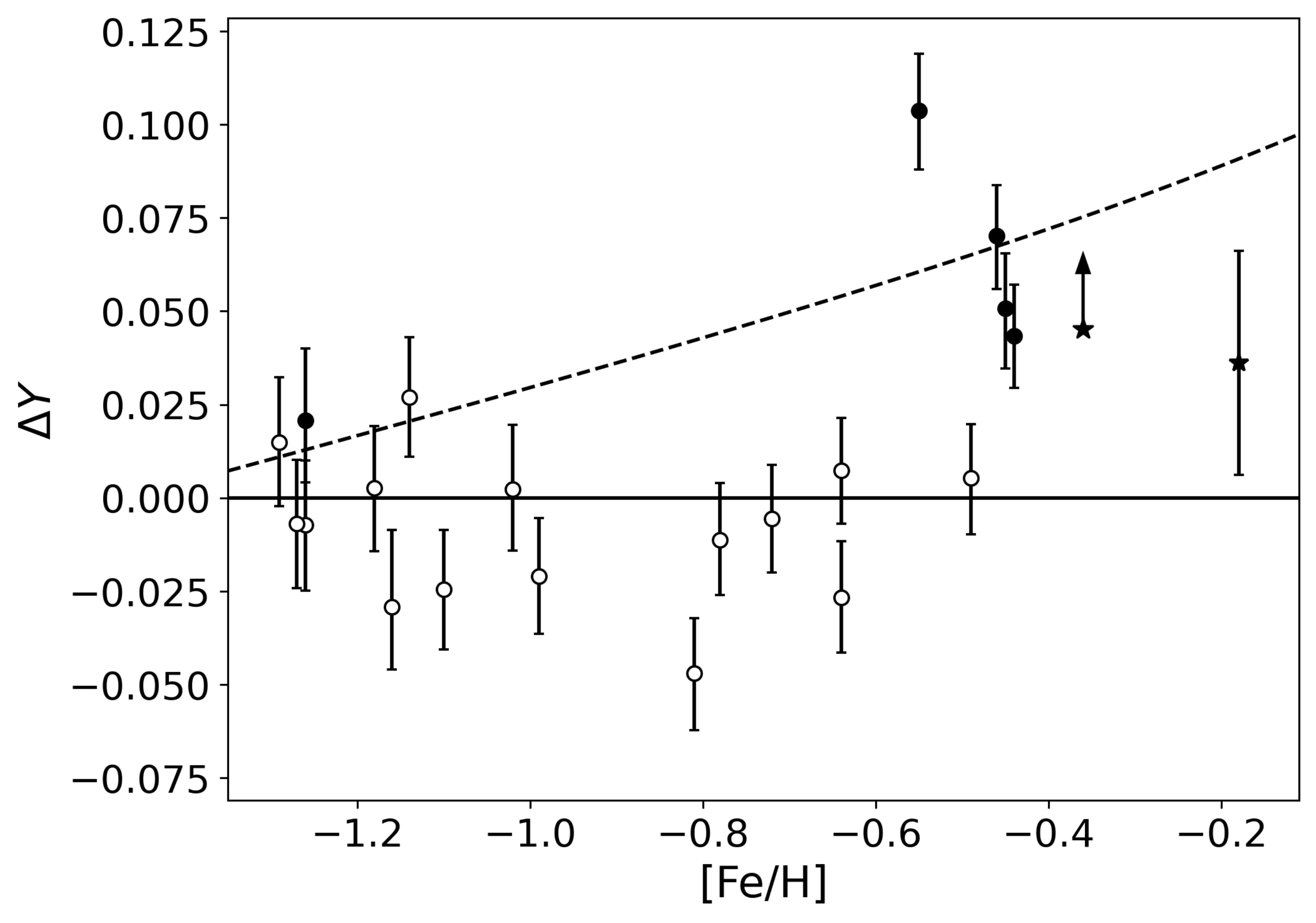}}
\caption{\label{fig:He2} The deviation in the Helium abundance from the Y--metallicity fit shown in Fig.\,\ref{fig:He1} against metallicity. The errors for each cluster are adjusted to account for the error in the fit and once again black markers correspond to upper branch candidates and white to lower. The dashed line shows the relationship $\Delta Y/\Delta Z = 1.54$ (Eq.\,\ref{eq:deltaY}). }
\end{figure}

We explore the impact of Helium abundances on the upper and lower branch splitting of H$\beta_o$.
Helium mass fraction data is provided by \citet{wagner-kaiser2017} under the assumption that GCs comprise single stellar populations. CMDs produced using photometry from \citet{sarajedini2007} were used to determine fits for isochrones of the GGCs.  The Bayesian analysis suite used, BASE-9, is a tool that can be employed for the fitting and characterization of GCs \citep[e.g.][]{stenning2016}. It uses adaptive Metropolis Markov chain Monte Carlo to estimate model parameters, mapping their full posterior distribution. This was done by sampling the joint posterior distribution of age, absorption, distance, metallicity and Helium of a cluster along with individual stellar parameters of binarity, zero-age main sequence mass and cluster membership. By using a wide variety of cluster and stellar parameters,  BASE-9 provides precise and reproducible fits which in turn give accurate and precise Helium mass fraction measurements \citep[for further details, see][]{wagner-kaiser2017}. 

We first considered the relationship between Helium abundance and metallicity. We found Helium abundance has a strong dependence on metallicity, which is followed strictly apart from a group of metal-rich clusters with uncharacteristically high Helium abundances, shown in Fig.\,\ref{fig:He1}. We remove the dependence on metallicity by fitting this relationship using a first order least squares polynomial. We then plot $\Delta Y$, the deviation from the fit for each cluster, against metallicity. This allows a comparison which is equivalent across all metallicities. This is shown in Fig.\,\ref{fig:He2} for intermediate to high metallicities ([Fe/H] $> -1.3$) with the Helium abundance data taken from \citet{wagner-kaiser2017}. We supplemented this data with two of our upper branch candidates; NGC\,6440 and NGC\,6553. The Helium abundance of NGC\,6440 assumption (Y $> 0.3$) is provided by \citet{mauro2012} who make this prediction based on the bimodality of its HB, stating they expect an "anomalously high" Helium abundance if the split in NGC\,6440 is due to evolutionary effects. They also suggest that the metal-rich cluster NGC\,6569, a lower branch candidate, is unlikely to have a high Helium content when interpreting these effects. This is further supported by \citet{Suk2008} who determine that high Helium abundance is one of the key drivers behind a bimodal HB most common in massive GCs. The Helium abundance of NGC\,6553 (Y = $0.28\pm 0.03$) is taken from \citet{Guarnieri1998} which is significantly above $\Delta Y = 0$. The error in this value is large compared to that in the rest of the data, but still it does not go below $\Delta Y = 0$. We are unable to find Helium abundance data for the upper branch candidates NGC\,6342 and NGC\,6528. 

From this plot we can see a clear relationship between the Helium abundance and upper branch selection. Of the seven upper branch GGCs we have data for, all are clearly above the $\Delta\,Y = 0$ line. We also see two lower branch candidates with high Helium abundance when taking into account their lower error limit; NGC\,5904 and NGC\,2808. Both
have \mgfep\,$< 1.8$\AA\ placing them in the lower metallicity section of the right panel of Fig.\,\ref{fig:HL} where their H$\beta_o$ measurements can be explained using Fig.\,\ref{fig:metal_poor}. Considering these results we determine that we are able to identify high Helium abundance clusters from enhanced H$\beta_o$ values at intermediate to high metallicities.
Taking this further, while the lower branch clusters follow the general trend, the upper branch clusters are well predicted by the slope of the model assumption used by \citet{Dotter2008}, $\Delta Y/\Delta Z = 1.54$, with an offset of $Y_o = 0.321\pm0.035$ (the peak of the Helium abundance MW distribution provided by \citealt{wagner-kaiser2017}). Agreeing well for all seven of our upper branch GGCs, the relation is defined by Eq.\,\ref{eq:deltaY}.

\begin{equation}\label{eq:deltaY}
    \Delta Y = (1.54\cdot Z + Y_o) - (-0.0564\cdot\log\left( \frac{Z}{0.0134}\right)\ + 0.24)
\end{equation}

Considering these results, we suggest that an enhanced Helium abundance is the cause of the splitting of H$\beta_o$ GGC measurements. Additionally, it is possible to identify high Helium abundance clusters from enhanced H$\beta_o$ values at intermediate to high metallicities using just integrated spectroscopy. \emph{Until now, no method for this existed in the literature.}

The determination of GC Helium content using resolved spectroscopy is only possible in rare cases, for only a small set of stars within the stellar population \citep[e.g.][]{Piotto2007, villanova2012, Marino2014}. However, we are unable to compare these measurements to the \citet{wagner-kaiser2017} data as each small set of stars will belong to one of the stellar populations within a cluster, whereas the \citet{wagner-kaiser2017} data provides an average Helium abundance considering each cluster as though they were composed of a single stellar population.

Having argued that an enhanced Helium abundance may be the cause of the increase in H$\beta_o$ for GC integrated spectra, we now explore why variations in Helium may affect the Hydrogen Balmer lines.
Ideally, we would report on the difference in H$\beta_o$ for standard and Helium enhanced SSP models. However, at present, there are no SSP models with empirical stellar libraries which self-consistently allow for varying Helium. Instead, we investigated how varying Y might affect Balmer-lines in integrated spectra, by exploring the impact of varying $Y$ on theoretical stellar spectra and on the isochrones (two key ingredients in SSP models).

\subsubsection{Helium enhanced synthetic stellar spectra}

To explore the impact of varying $Y$ on the Balmer lines (and specifically, the low resolution Balmer indices) in typical GC stars, synthetic stellar spectra were generated. Stars were created with parameters similar to those expected for stars around the MSTO in the moderately metal-rich cluster 47\,Tuc (NGC\,104), as they contribute the most to the strength of the Balmer lines of the integrated light of 47\,Tuc \citep{vazdekis2001}. For this purpose, stars were produced with \Feh$=-0.76, T_\text{eff}\,=\,5000,\,5500,\,6000\text{K}, \log g = 4.0, 4.5$ and $Y=0.25, 0.35$. The values for Helium were selected to reflect the mean differences in $Y$ between upper and lower branch clusters seen in Fig\,\ref{fig:He1}. The models maintained a fixed microturbulence, $v_\text{turb}=2.0$\,kms$^{-1}$ and stellar atmospheric models were generated with the specified stellar parameters. Then, a synthetic spectrum was created for the star with the same stellar parameters. For the synthetic spectral computation of \textsc{ATLAS9} models, written by \citet{Kurucz1979, Kurucz1993}, \textsc{SYNTHE} was used with atomic and molecular linelists from the Castelli website (\url{http://wwwuser.oats.inaf.it/castelli/}). These lists were originally compiled by \citet{kurucz1991} and later updated by \citet{castelli2004}. The \textsc{ATLAS9} code computes plane-parallel hydrostatic models in Local Thermodynamical Equilibrium. The code allows arbitrary chemical compositions, and a collection of Opacity Distribution Functions for various metallicities are available to account for the effect of line opacity \citep[ISPy3,][]{ISPy3}.

The models were produced at $R\sim500,000$ and degraded to 3.1\,\AA\ FWHM to allow for direct comparison with our combined GGC dataset and in order to measure their line-strength indices. We focus our analysis on models with $T_\text{eff}=6000\,\text{K}$ with Helium $Y_1=0.25$ and $Y_2=0.35$ as well as $\log g = 4.0$ and $4.5$. The difference in H$\beta_o$ between the two Helium values for these MSTO stars provides an upper limit for the H$\beta_o$ variation in the spectrum of the SSP for such Helium enhancement. The H$\beta_o$ difference for the models with $\log g = 4.0$ and $4.5$ are both $0.117$\AA; the change in stellar gravity has no influence on the relative H$\beta_o$ difference at this temperature and metallicity regime. A second set of models were produced by (C. Allende Prieto, private comm.) in a similar fashion to ours and are mostly equivalent apart from that the increase in Helium from the base models was $\sim50\%$ less: $Y \sim 0.3$ \citep[see e.g.][for the description of similarly produced model spectra]{Carlos2018, SMILES}. The change in H$\beta_o$ for both of the gravity values are $0.055$\AA, which is $\sim50\%$ less than the change in H$\beta_o$ for our models, suggesting that the effects of Helium on H$\beta_o$ are approximately linear between at least $Y = 0.25$ and $0.35$.

\begin{figure}
\centerline{\includegraphics[width=.94\columnwidth, left]{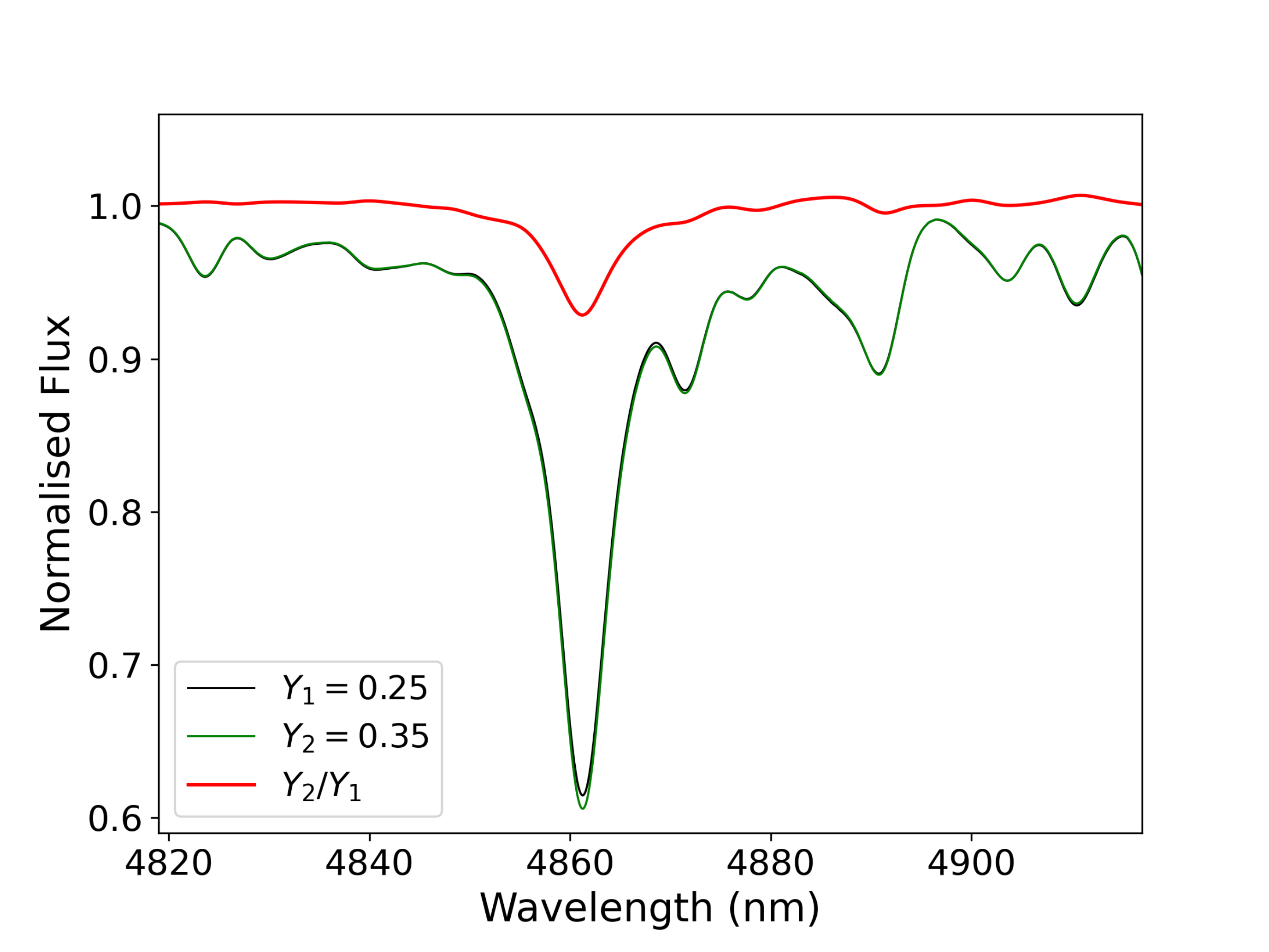}}
\caption{\label{fig:Y2vsY1}Synthetic stellar spectra with $Y_1 = 0.25$ (black line) and $Y_2 = 0.35$ (green line) in the wavelength range of the H$\beta$ spectral feature index definition, smoothed to a resolution of $3.1$\,\AA\ FWHM. The red line shows the ratio $Y_2/Y_1$ which has been multiplied by a factor of 5 and re-centered around 1 for clarity.}  
\end{figure}

\begin{figure}
\centerline{\includegraphics[width=.94\columnwidth, left]{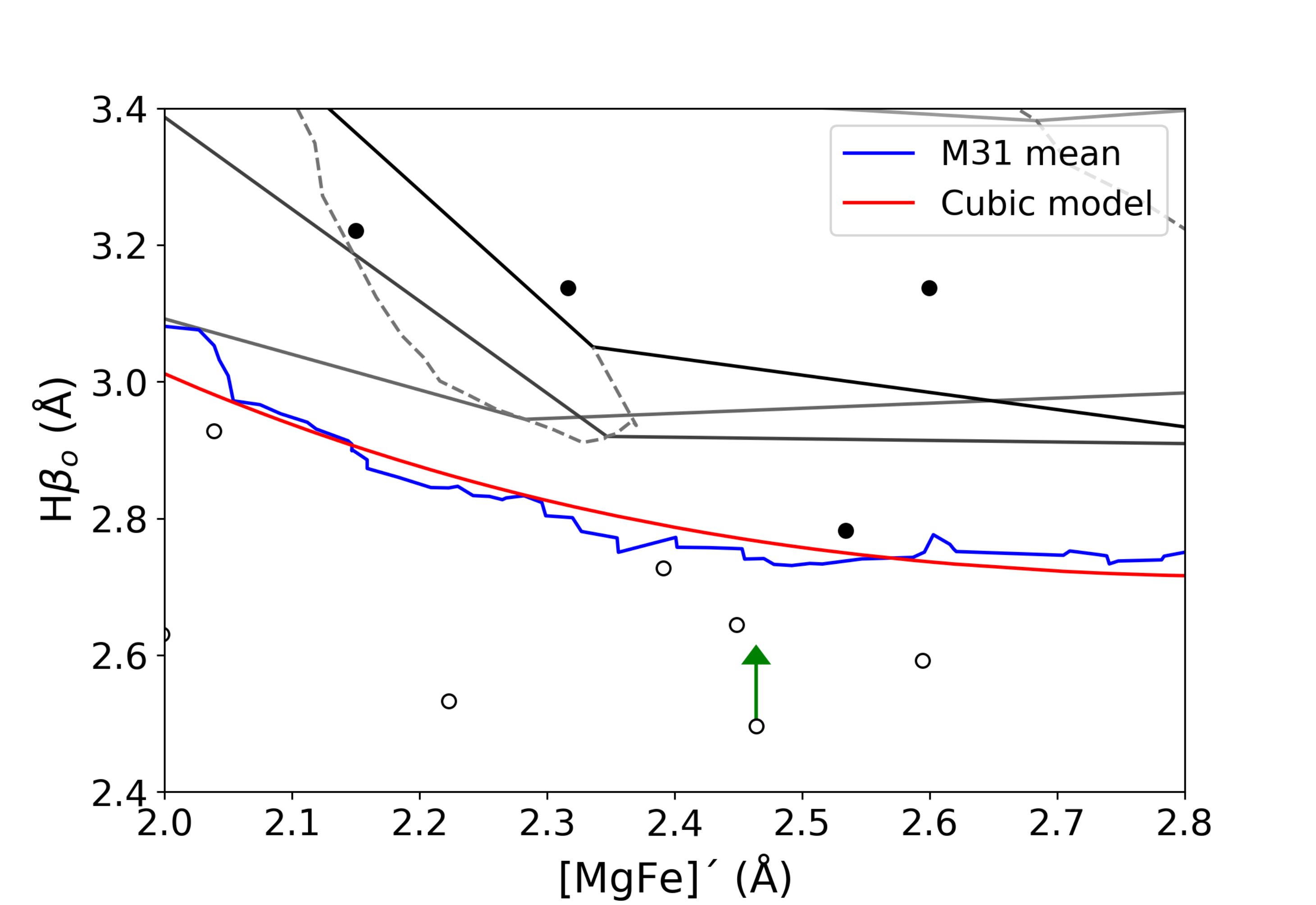}}
\caption{\label{fig:HeM}An enlarged version of the left panel of Fig.\,\ref{fig:HL} with \mgfep\ against H$\beta_o$. The green arrow represents the H$\beta_o$ change due to enhancing Helium from $Y = 0.25$ to $Y = 0.35$ for stars resembling the MSTO of 47\,Tuc. The running mean and its fit are shown by the blue and red line and separate the upper and lower branch.}  
\end{figure}

In Fig.\,\ref{fig:Y2vsY1}, we compare the spectra around the H$\beta$ spectral feature with the black line showing the base helium spectra and the green line the helium enhanced. The red line represents the ratio between the spectra ($Y_2/Y_1$) and highlights their differences. The shape of the ratio between the spectra at wavelengths equivalent to the H$\beta$ feature demonstrate an increase in depth and width of the spectral feature for the $Y_2$ spectra for our models. It follows that the $Y_2$ spectrum has a higher H$\beta_o$ measurement.
Our models are shown in Fig.\,\ref{fig:HeM} where the green arrows correspond to the change in H$\beta_o$ for an increase in Helium of $Y = 0.25$ to $Y = 0.35$. As the modelled stars are supposed to mimic those of 47\,Tuc, the arrow begins at the H$\beta_o$ and \mgfep\ of 47\,Tuc. A shift equal to the model arrow is not significant enough to take 47\,Tuc into the upper branch (above the red or blue line). Also, it is worth noting that 47\,Tuc has a Helium abundance of $Y = 0.271$ provided by \citet{wagner-kaiser2017}. Therefore, the simulated base level of Helium, $Y=0.25$, is lower than the true level and we would expect the increase in H$\beta_o$ for the hypothetical, Helium enhanced ($Y=0.35$), 47\,Tuc to be lower. However, we must note that our models do suggest a significant change in H$\beta_o$ which is in agreement with our prior conclusion with respect to the positive direction of the change. Furthermore, the change in H$\beta_o$ is significant and, if applied to other GCs at similar metallicities (NGC\,6316, NGC\,6637), would result in the satisfaction of Eq.\,\ref{eq:model}. Now, our theoretical stellar spectra models need to be coupled with an understanding of Helium enhanced isochrones to estimate what the overall effect of Helium enhancement has on the H$\beta_o$ measurement of SSPs. 

\subsubsection{Helium enhanced isochrones}

Isochrones of varying Helium were produced by \citet{valcarce2012} and are inclusive of the range explored by our Helium enhanced stellar spectra. Particular attention should be paid to Fig.~8 in \citet{valcarce2012} which shows the $\log(T_\text{eff})$ against $\log(\text{L}/\text{L}_\odot)$ of isochrones with three metallicities and ranging from $Y=0.245$ to $Y=0.370$, all for four ages (7.5~Gyr, 10~Gyr, 12.5~Gyr, 15~Gyr). The isochrones closest in \Feh\ and age to our enhanced stellar spectra are those labelled $Z_2$ ($\Feh \sim -0.92$) for 12.5~Gyr (10~Gyr isochrones are also relevant for the old GC population). The turn-off for this set of isochrones changes position with $Y$, where as Helium abundance increases so does the effective temperature of the turn-off. Due to the Balmer lines sensitivity to the turn-off $T_\text{eff}$, these isochrones suggest an increase in Helium in a SSP results in an increase in H$\beta_o$. If these isochrones were combined with Helium enhanced stellar spectral libraries to produce Helium enhanced SSPs, the evidence suggest that both ingredients would cause an increase in H$\beta_o$. While we cannot give a quantitative estimate for the H$\beta_o$ difference caused by an increase in Helium (representative of the difference in Helium for upper and lower branch cluster), we can say that at intermediate to high metallicities a) an increase in Helium results in an increase in H$\beta_o$ and b) the increase is significant with respect to a GCs upper or lower branch classification. Both lend weight to our previous argument that the upper and lower branch splitting of GCs at intermediate to high metallicities is caused by differences in Helium abundance. We note that \citet{vazdekis2001} explored the impact of varying Helium on the index H$\gamma_{\sigma < 130}$ using SSP models where the integrated spectra (fixed Helium) were synthesized on the basis of different isochrones (varying Helium). However, due to their consideration of a different index and small Helium variations compared to those we are concerned with, we do not consider their results further here. 

\subsubsection{The relation of Helium to further cluster parameters}

\begin{figure}
\centerline{\includegraphics[width=.94\columnwidth, left]{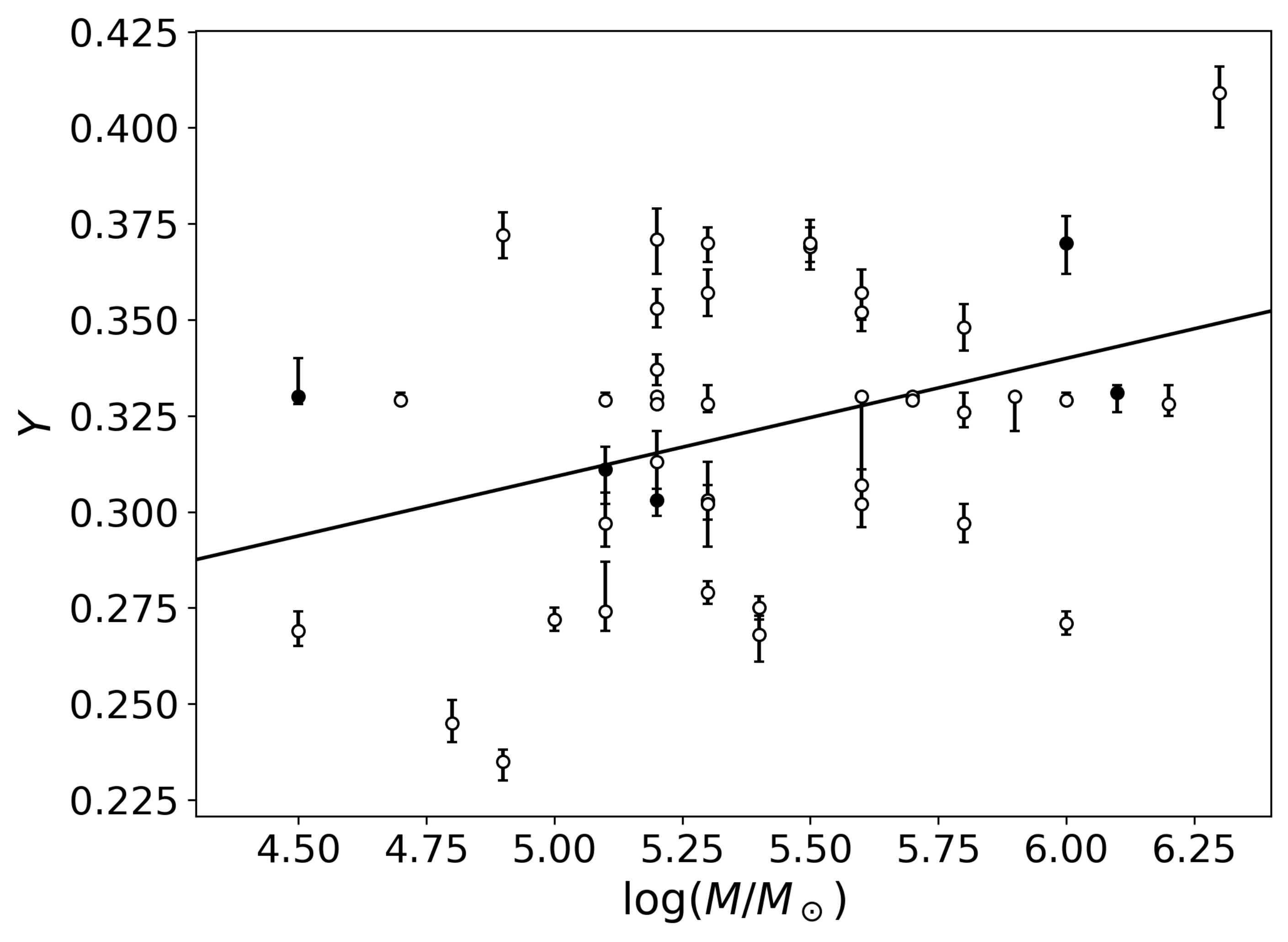}}
\caption{\label{fig:YvMass}The Helium mass fraction against the logarithm of the cluster mass in stellar masses. A positive correlation suggests that more massive clusters show greater Helium enrichment. The first order least squares polynomial (solid black line) is of the form $Y = (0.0308\pm0.0125)\log(\mathrm{M/M}_{\odot}) + (0.155\pm0.068)$. Black markers correspond to upper branch candidates and white markers lower. }
\end{figure}

\begin{figure}
\centerline{\includegraphics[width=.94\columnwidth, left]{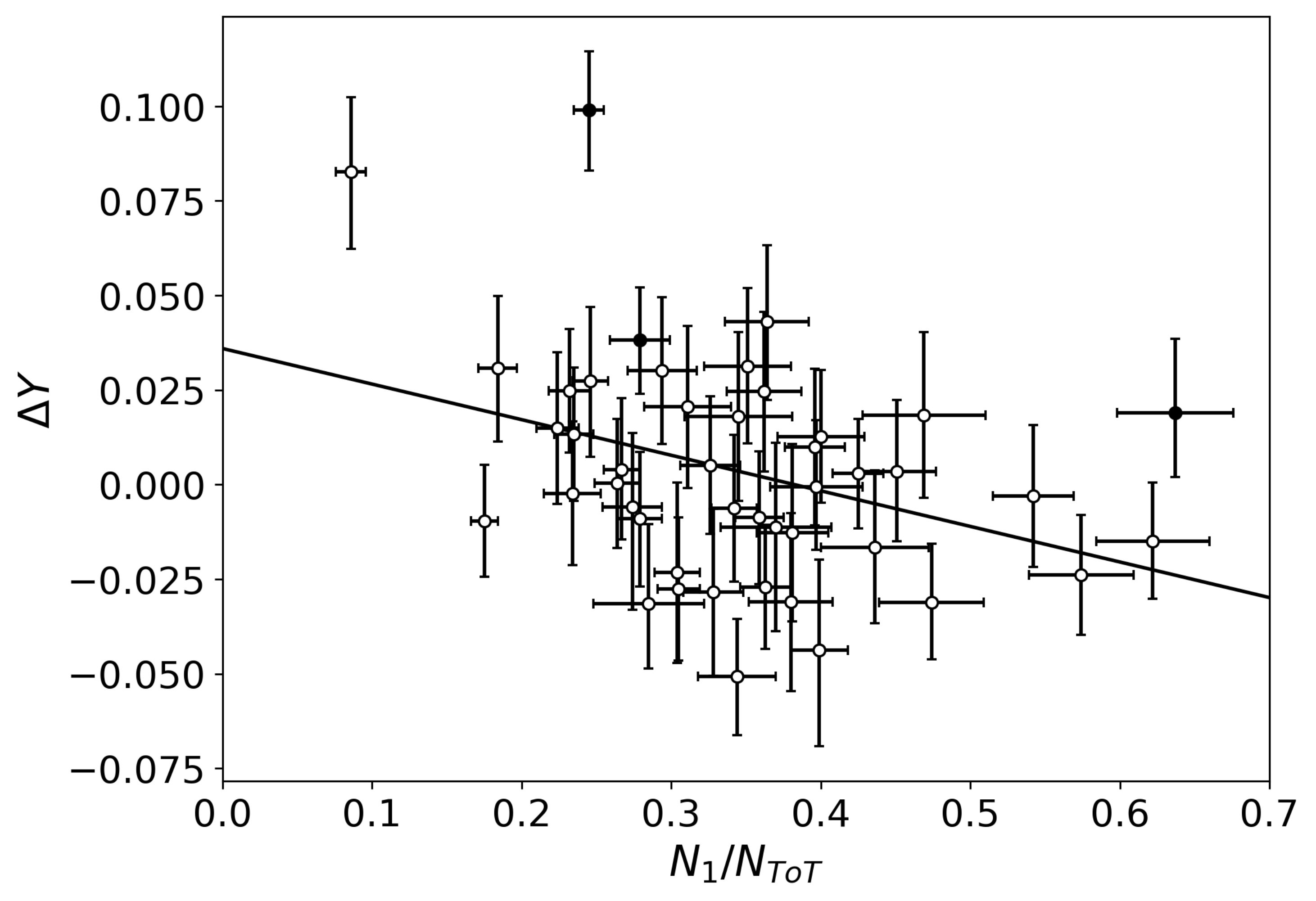}}
\caption{\label{fig:Yvmp} The metallicity independent Helium abundance, $\Delta\,Y$, against the fraction of first generation stars of each GGC. The first order least squares polynomial (solid black line) is of the form $\Delta\,Y = (-0.0941\pm0.0373) N_1/N_{ToT} + (0.0360\pm0.0135)$. Black markers correspond to upper branch candidates and white markers lower. }
\end{figure}

Previously, Helium abundance has been shown to correlate with cluster magnitude and binary fraction \citep[see][]{wagner-kaiser2017}. The absolute magnitude of a cluster can be used as a proxy for its mass and more massive clusters are expected to have a higher binary fraction \citep{milone2012acs}. Here, we present the direct comparison of cluster mass and Helium abundance. The cluster mass data is taken from \citet{waggs} and have been provided by varying sources \citep[e.g.][]{Harris1996, McLaughlin2005}. Shown in Fig.\,\ref{fig:YvMass}, we find the expected correlation with higher cluster mass corresponding to greater Helium enhancement but only with a moderate Pearson's correlation coefficient of $0.351\pm0.143$. However, we are unable to identify a difference between the upper and lower branches cluster mass or ratio of stellar populations. This is shown by the lack of a clear difference in the relationship for black (upper branch) and white (lower branch) markers.
In \citet{Milone2017}, the fraction of first generation stars was shown to anti-correlate with cluster mass and \citet{lucatello2015} showed that the more Helium poor first generation stars have a higher binary fraction compared to second generation stars. However, the results of the recent study by \citet{milone2020} suggest equal binary fractions for the majority of clusters. Also, the internal Helium abundance varies significantly between first and second generation stars for some clusters \citep{Milone2018}. These studies combined with our results suggest an anti-correlation between Helium abundance and the ratio of multiple stellar populations (fraction of first generation stars). As expected, we identify a direct anti-correlation between the Helium abundance of a cluster and the fraction of first generation stars provided by \citet{Milone2017}, suggesting later generation stars are more Helium enhanced than their first generation counterparts. However, once again we do not see a difference in the relationship between upper and lower branch clusters and the fraction of first generation stars. We now consider the relationship between the metallicity insensitive $\Delta\,Y$ and the fraction of first generation stars. This is shown in Fig.\,\ref{fig:Yvmp} and has a Pearson's correlation coefficent of $-\,0.360\pm0.142$. Disappointingly data is only available for three of seven upper branch clusters; NGC\,6388, NGC\,6624 and NGC\,6717. Attempts were made to disentangle the first and second generations stars in upper branch clusters NGC\,6304, and NGC\,6441 but failed, requiring greater photometric accuracy \citep{Milone2017}. The three upper branch clusters with data available do show the beginnings of a different relationship for upper branch clusters with NGC\,6388 and NGC\,6717 above the main the main locus and NGC\,6624 at its tip. However, with just three measurements, we cannot draw any convincing conclusions. We are also unable to identify a significant relationship between the Helium variation between first and second generations stars and $Y$, $\Delta\,Y$ or upper branch candidacy.

In the following section we illustrate the potential use of the results obtained here for predicting the Helium abundance in GCs of external galaxies.

\subsection{\label{sec:m31}M31 clusters}

We now look to use the relation given by Eq.\,\ref{eq:deltaY} to predict Helium abundances for upper branch M31 clusters highlighted in the right panel of Fig.\,\ref{fig:HL}: B129, B082-G144 and B030-G091. These clusters are used as an example to demonstrate the methodology and are of the metallicity $-0.8 \pm 0.1$, $-0.7 \pm 0.1$ and $-0.3 \pm 0.1$ respectively, provided by \citet{Nelson2011}. Using this and their upper branch status we predict $\Delta Y$ values of $0.039 \pm 0.038$, $0.046 \pm 0.038$ and $0.074 \pm 0.038$ using the $\Delta Y/\Delta Z = 1.54$ relationship. Errors have been calculated using a combination of the error in metallicity, the slope and intercept of the fit and the peak Helium abundance value as in Fig.\,\ref{fig:He2}. Using these values we also calculate absolute Helium abundances of $Y =  0.324 \pm 0.035$, $Y =  0.325 \pm 0.035$ and $Y = 0.331 \pm 0.035$ respectively. 

We are unable to provide reliable estimates for cluster mass and the ratio of stellar populations due to the lack of a difference of relationship for the upper and lower branch for these parameters. However, the relationship demonstrated in  Fig.\,\ref{fig:YvMass} and the high Helium abundances of the three upper branch clusters suggest that these cluster will have comparatively high cluster masses. Two of the three clusters (B082-G144 and B030-G091) have virial mass estimates, provided by \citet{Strader2011}, of $\log\text{M}_v = 6.52^{+0.005}_{-0.005}$ and $5.39^{+0.006}_{-0.007}\,\text{M}_\odot$ respectively. The spread between these two clusters is considerable, where by consulting Fig.\,\ref{fig:YvMass} it can be seen that B030-G091 has a moderate mass while B082-G144 has a large mass, to the high end of the plot when compared to MW GCs. This somewhat disagrees with our prior statement that we would expect higher cluster masses for these three GCs and highlights the limits of predicting upper branch cluster masses, regardless of their Helium enhanced status. We are unable to use Fig.\,\ref{fig:Yvmp} to provide solid predictions for the ratio of first generation stars for our clusters due to the aforementioned lack of upper branch measurements available. A major assumption we have made while making these predictions is that the clusters of M31 have the same relationship between Helium and metallicity as the GGCs. We are aware that M31 clusters show a broad, unimodal metallicity distribution in contrast to the well established bimodal distribution in the MW, suggesting a different star formation and accretion history \citep{Nelson2011, cezario2013}. Even though this could effect the validity of our absolute Helium abundance predictions, it should not effect their status as being Helium enhanced (this applies to all M31 upper branch clusters). 

\citet{Nelson2011} identified six M\,31 EGCs using integrated spectroscopic methods, having intermediate ages of around $7$\,Gyr: B015, B071, B138, B140, B268, and AU010. All of these EGCs are metal-rich where five out of six have \Feh$\ > -0.2$\,dex, matching the two upper branch GGCs NGC\,6528 and NGC\,6553. Five out of six of the intermediate age EGCs have mass $5.0\,<\,\log(\text{M/M}_\odot)\,\lesssim\,5.6$ which is shown to be average by Fig.\,\ref{fig:YvMass}. Due to the loose correlation between cluster mass and Helium abundance, this suggests an average Helium abundance. However, this does not take into account the EGC's metallicity or correlate to lower branch candidacy. Considering the lack of $\sim 7$\,Gyr GGCs at high metallicities, we suggest that the ages of these clusters were possibly underestimated due to anomalously high Helium abundance at these metallicities, and they are in fact old ($\ge 10$\,Gyr). However, due to the aforementioned evidence supporting differences in the history of M\,31 and MW clusters, it is also possible that these are in fact intermediate age clusters and exist in contradiction to the GGCs observed at these metallicities.

\section{Summary and conclusions}

We aimed to identify the cause of the enhancement of GGC Balmer spectral line-strength measurements which can result in the under-estimation of spectroscopic ages.

The best possible analysis with the current GGC data available required the homogenisation of three separate datasets (WAGGS, \citetalias{Kim2016} and \citetalias{S05}) for \mgfep\ and H$\beta_o$ measurements, while considering their radial dependence. H$\beta_o$ exhibited a significant radial dependence and, therefore, we produced models to simulate the effect of mass segregation on H$\beta_o$. We found they did not match observations, suggesting mass segregation plays a minimal role in the radial dependence of H$\beta_o$.

In contention with \citetalias{cenarro2008}, we only identify an upper and lower branch at intermediate to high metallicities (\Feh$>\,-1.3$\,dex, \mgfep\,$> 1.8$\,\AA) and find no correlation between upper branch status and the specific fraction of BSSs; previous assumptions were driven by the splitting of H$\beta_o$ at lower metallicites that is no longer observed. Also, we provide a definition of upper lower branch in terms of \mgfep\ and H$\beta_o$: \mgfep\,$> 1.8$\AA\ (\Feh$ > 1$) and Eq.\,\ref{eq:model}. Our analysis was limited by the lack of a larger BSS dataset produced using modern techniques when compared to the data used in \citetalias{cenarro2008}. Modern catalogues are restricted by observational issues that arise from searching for BSSs in the optical passband. Optical emission in GCs is heavily influenced by red (cool), bright RGB and SGB stars compared to the hot but faint BSSs. The solution to this problem is to search for BSSs in the UV, where RGB and SGB stars are relatively faint compared to the bright BSSs \citep[see][]{Raso2017}. 

We conclude that an enhanced Helium abundance is the primary cause of observed splitting of H$\beta_o$ at intermediate to high metallicities. To explore this further, we produced a metallicity-insensitive Helium abundance and investigated its relationship with upper branch candidacy. We found that their relationship is well described by the slope of the \citet{Dotter2008} model prediction $\Delta Y/\Delta Z = 1.54$ with an offset of $Y_o = 0.321\pm0.035$. The quality of this fit is limited by the small number of upper branch GGCs present in the GGC sample, bringing into question the absolute Helium abundance predictions but not effecting their Helium enhanced status. The consequence of enhanced Helium with relation to Balmer line strength was further explored by examining the change in H$\beta_o$ with increased Helium abundance for two ingredients of SSPs: stellar spectra and isochrones. Modelled stellar spectra with varying Helium for stars with parameters reflecting those near the MSTO of 47\,Tuc suggest an increase in Helium results in increased H$\beta_o$, while Helium enhanced isochrones also suggest an increase in H$\beta_o$, but would benefit from further exploration.

We compared the Helium abundances of our GGC sample to the fraction of first generation stars, $N_1/N_{ToT}$, which gives the ratio of multiple stellar populations. Interestingly, we find that the two upper branch clusters with both Helium and $N_1/N_{ToT}$ measurements are separate from the main locus of lower branch clusters; suggesting the ratio of multiple stellar populations can be inferred from integrated spectroscopy. These different generations have differing chemical abundance patterns (e.g. oxygen, sodium), possibly allowing their inference from integrated spectroscopy using this methodology. Also, we report a loose correlation between GGC Helium abundance and cluster mass; important in light of the multiple stellar population scenario as an increase in cluster mass is known to increase both its incidence and complexity.

The methodology developed via the study of GGCs was then used to explore the M\,31 EGC population. This new methodology allows for the inference of the Helium abundances for EGCs using just integrated spectroscopy, the first of its kind in the literature. Subsequently, we predict the Helium abundance of three M\,31 EGCs: B129 ($Y =  0.324\pm0.035$), B082-G144 ($Y =  0.325\pm0.035$) and B030-G091 ($Y = 0.331\pm0.035$). By the application of this methodology, Helium abundance predictions could be made for the whole sample of M\,31 GCs at intermediate to high metallicities. 

In addition, we question the intermediate spectroscopic age measurements ($\sim 7$\,Gyr) of \citet{Nelson2011} for six M\,31 EGCs: B015, B071, B138, B140, B268, and AU010. We suggest that these six clusters may instead be old ($\ge 10$\,Gyr) with enhanced Helium abundance.

Based on the results of this study, we propose several avenues of future work. To identify the cause of the radial dependence of H$\beta_o$, two alternative causes should be investigated; stochastic effects and the radial distribution of multiple stellar populations. Also, to better assess the role of BSSs in the splitting of H$\beta_o$, a large, homogeneous UV BSSs sample, such as the upcoming catalogue eluded to in \citet{ferraro2018}, is required. Finally, in order to explore the possibility of the inference of multiple stellar populations from just GC integrated spectra, further work should focus on disentangling the multiple stellar populations of more upper branch clusters with an aim to provide their $N_1/N_{ToT}$ values.

The results of this study allow for the first predictions of Helium abundance and more accurate predictions of age for EGCs. The new methods presented in this paper can be applied to EGCs from a whole host of galaxies where integrated spectra are available; allowing the accurate prediction of age and Helium abundance, resulting in an increased understanding of the dynamical and chemical nature of the EGC population. In turn, the origins and chemo-dynamical evolution of galaxies outside of our own can be further understood.

\label{sec:sum}

\section{acknowledgements}

We thank Nelson Caldwell for kindly providing the M31 GC spectroscopic data and Carlos Allende Prieto for providing us with a second set of model stellar spectra to explore the impact of Helium, as well as informative discussions surrounding discrepancies when compared to our models.
H.J.L. acknowledges the support of the ERASMUS+ programme in the form of a traineeship grant and that this work has partially been carried out within the framework of the National Centre for Competence in Research (NCCR) PlanetS supported by the Swiss National Science Foundation (SNSF). 
M.A.B., A.V. and N.S.R. acknowledge financial support comes from the grant PID2019-107427GB-C32 from the Spanish Ministry of Science, Innovation and Universities (MCIU). M.A.B. acknowledges financial support from  the  Severo  Ochoa  Excellence  scheme(SEV-2015-0548). This work was backed through the IAC project TRACES which is partially supported through the state budget and the regional budget of the Consejería de Economía, Industria, Comercio y Conocimiento of the Canary Islands Autonomous Community

\section*{Data Availability}

The E-MILES SSP models are publicly available at the MILES website (\url{http://miles.iac.es}). 
The Kim et al. 2016 spectroscopic data are also available on the MILES website under "other predictions/data".
The WAGGS spectroscopic data are publicly available at (\url{https://www.astro.ljmu.ac.uk/~astcushe/waggs/data.html}). The Schiavon et al. 2005 spectroscopic data are publicly  available at (\url{https://www.noao.edu/ggclib/}).



\bibliographystyle{mnras}
\bibliography{mnras_template} 





\bsp	
\label{lastpage}
\end{document}